\documentclass[draft]{agujournal2019}
\usepackage{url} %this package should fix any errors with URLs in refs.
\usepackage{lineno}
\usepackage[inline]{trackchanges} %for better track changes. finalnew option will compile document with changes incorporated.
\usepackage{siunitx}
\usepackage{soul}
\usepackage{amsmath}
\usepackage{amsfonts}       % blackboard math symbols
\usepackage{pdflscape}
\usepackage[inkscapelatex=false]{svg}

\draftfalse

\journalname{AGU Advances}

\begin{document}

%%%%%%%%%%%%%%%%%%%%%%%%%%%%%%%%%%%%%%%%%%%%%%%
%  TITLE
%
% (A title should be specific, informative, and brief. Use
% abbreviations only if they are defined in the abstract. Titles that
% start with general keywords then specific terms are optimized in
% searches)
%
%%%%%%%%%%%%%%%%%%%%%%%%%%%%%%%%%%%%%%%%%%%%%%%

% Example: \title{This is a test title}

\title{HiRO-ACE: Fast and skillful AI emulation and downscaling trained on a 3 km global storm-resolving model}

%%%%%%%%%%%%%%%%%%%%%%%%%%%%%%%%%%%%%%%%%%%%%%%
%
%  AUTHORS AND AFFILIATIONS
%
%%%%%%%%%%%%%%%%%%%%%%%%%%%%%%%%%%%%%%%%%%%%%%%

% Authors are individuals who have significantly contributed to the
% research and preparation of the article. Group authors are allowed, if
% each author in the group is separately identified in an appendix.)

% List authors by first name or initial followed by last name and
% separated by commas. Use \affil{} to number affiliations, and
% \thanks{} for author notes.
% Additional author notes should be indicated with \thanks{} (for
% example, for current addresses).

% Example: \authors{A. B. Author\affil{1}\thanks{Current address, Antartica}, B. C. Author\affil{2,3}, and D. E.
% Author\affil{3,4}\thanks{Also funded by Monsanto.}}

\authors{W. Andre Perkins\affil{1}, Anna Kwa\affil{1}, Jeremy McGibbon\affil{1}, Troy Arcomano\affil{1}, Spencer K. Clark\affil{1,2}, Oliver Watt-Meyer\affil{1}, Christopher S. Bretherton\affil{1}, Lucas M. Harris\affil{2}}

% \affiliation{1}{First Affiliation}
% \affiliation{2}{Second Affiliation}
% \affiliation{3}{Third Affiliation}
% \affiliation{4}{Fourth Affiliation}

\affiliation{1}{Allen Institute for Artificial Intelligence, Seattle, WA, USA}
\affiliation{2}{NOAA/Geophysical Fluid Dynamics Laboratory, Princeton, NJ, USA}
%(repeat as many times as is necessary)

% Corresponding author mailing address and e-mail address:

% (include name and email addresses of the corresponding author.  More
% than one corresponding author is allowed in this LaTeX file and for
% publication; but only one corresponding author is allowed in our
% editorial system.)

% Example: \correspondingauthor{First and Last Name}{email@address.edu}

\correspondingauthor{W. Andre Perkins}{andrep@allenai.org}

%%%%%%%%%%%%%%%%%%%%%%%%%%%%%%%%%%%%%%%%%%%%%%%
% KEY POINTS
%%%%%%%%%%%%%%%%%%%%%%%%%%%%%%%%%%%%%%%%%%%%%%%
%  List up to three key points (at least one is required)
%  Key Points summarize the main points and conclusions of the article
%  Each must be 140 characters or fewer with no special characters or punctuation and must be complete sentences

% Example:
% \begin{keypoints}
% \item	List up to three key points (at least one is required)
% \item	Key Points summarize the main points and conclusions of the article
% \item	Each must be 140 characters or fewer with no special characters or punctuation and must be complete sentences
% \end{keypoints}

\begin{keypoints}
\item HiRO-ACE skillfully emulates mean and extreme precipitation from a `digital twin' 10 year 3-km grid X-SHiELD global atmospheric simulation
\item HiRO perfect-prediction diffusion-based downscaling is matched with ACE2S, a stochastic version of the ACE2 coarse model emulator
\item HiRO-ACE generates near-global 3 km precipitation fields two orders of magnitude more efficiently than X-SHiELD
%\item HiRO-ACE allows for seamless global 3 km downscaling of surface precipitation from a coarse climate emulator
\end{keypoints}

%%%%%%%%%%%%%%%%%%%%%%%%%%%%%%%%%%%%%%%%%%%%%%%
%
%  ABSTRACT and PLAIN LANGUAGE SUMMARY
%
% A good Abstract will begin with a short description of the problem
% being addressed, briefly describe the new data or analyses, then
% briefly states the main conclusion(s) and how they are supported and
% uncertainties.

% The Plain Language Summary should be written for a broad audience,
% including journalists and the science-interested public, that will not have 
% a background in your field.
%
% A Plain Language Summary is required in GRL, JGR: Planets, JGR: Biogeosciences,
% JGR: Oceans, G-Cubed, Reviews of Geophysics, and JAMES.
% see http://sharingscience.agu.org/creating-plain-language-summary/)
%
%%%%%%%%%%%%%%%%%%%%%%%%%%%%%%%%%%%%%%%%%%%%%%%

%% \begin{abstract} starts the second page

\begin{abstract}
Kilometer-scale simulations of the atmosphere are an important tool for assessing local weather extremes and climate impacts, but computational expense limits their use to small regions, short periods, and limited ensembles.  Machine learning offers a pathway to efficiently emulate these high-resolution simulations.  Here we introduce HiRO-ACE, a two-stage AI modeling framework combining a stochastic version of the Ai2 Climate Emulator (ACE2S) with diffusion-based downscaling (HiRO) to generate 3 km precipitation fields over arbitrary regions of the globe. Both components are trained on data derived from a decade of atmospheric simulation by X-SHiELD, a 3 km global storm-resolving model.  HiRO performs a 32x downscaling---generating 3 km 6-hourly precipitation from coarse 100 km inputs by training on paired high-resolution and coarsened X-SHiELD outputs. ACE2S is a $1^\circ \times 1^\circ$ ($\sim$100 km) stochastic autoregressive global atmosphere emulator that maintains grid-scale precipitation variability consistent with coarsened X-SHiELD, enabling its outputs to be ingested by HiRO without additional tuning. HiRO-ACE reproduces the distribution of extreme precipitation rates through the 99.99th percentile, with time-mean precipitation biases below 10\% almost everywhere. The framework generates plausible tropical cyclones, fronts, and convective events from poorly resolved coarse inputs. Its computational efficiency allows generation of 6-hourly high-resolution regional precipitation for decades of simulated climate within a single day using one H100 GPU, while the probabilistic design enables ensemble generation for quantifying uncertainty. This establishes an AI-enabled pathway for affordably leveraging the realism of expensive km-scale simulations to support local climate adaptation planning and extreme event risk assessment.

\end{abstract}

\section*{Plain Language Summary}
Climate models that simulate weather at fine scales (around 3 kilometers) can capture important details about heavy rainfall events from thunderstorms and tropical cyclones that coarser models fail to reproduce. However, these high-resolution simulations require massive computing power, limiting how often scientists can run them and making it difficult to explore the range of impacts in potential climates. We developed HiRO-ACE, an end-to-end AI system that learns to mimic an expensive high-resolution climate model from nearly ten years of simulation output. It works in two stages: first, a fast model generates global weather patterns at coarse resolution, then a second model fills in fine-scale precipitation details for a specified region. HiRO-ACE reproduces the frequency of extreme rainfall events and generates realistic storm features that only appear at fine scales. The system runs hundreds of times faster than traditional simulations, enabling researchers to generate many possible climate outcomes efficiently. This capability is a step towards helping communities better prepare for future climate risks by providing an accessible method to make local projections of extreme weather that were previously too expensive to produce.

%%%%%%%%%%%%%%%%%%%%%%%%%%%%%%%%%%%%%%%%%%%%%%%
%
%  BODY TEXT
%
%%%%%%%%%%%%%%%%%%%%%%%%%%%%%%%%%%%%%%%%%%%%%%%

%%% Suggested section heads:
% \section{Introduction}
%
% The main text should start with an introduction. Except for short
% manuscripts (such as comments and replies), the text should be divided
% into sections, each with its own heading.

% Headings should be sentence fragments and do not begin with a
% lowercase letter or number. Examples of good headings are:

% \section{Materials and Methods}
% Here is text on Materials and Methods.
%
% \subsection{A descriptive heading about methods}
% More about Methods.
%
% \section{Data} (Or section title might be a descriptive heading about data)
%
% \section{Results} (Or section title might be a descriptive heading about the
% results)
%
% \section{Conclusions}

\section{Introduction}

Kilometer-scale climate modeling provides essential capabilities for investigating local environmental impacts under a changing climate. Models of earth system components such as the atmosphere, land, and ocean with horizontal grid spacings of 1-5~km can have higher fidelity since they explicitly resolve key physical processes for which coarser-grid models use subjective, uncertain parameterizations.  For instance, atmospheric models at this scale can resolve important mesoscale features of fronts, tropical cyclones, thunderstorms, and terrain interactions, which leads to a more accurate prediction of precipitation over land \cite{Bretherton2022} and gives the visual impression of a ``digital twin" of Earth's atmosphere \cite{Ste2019,Cheng2022,DoblasReyes2025,Klocke2025}. However, their extreme computational resource requirements of processing and storage present a significant barrier to wider use and adoption by the climate science and adaptation communities. In particular, large ensembles of multi-decadal simulations for generating robust climate statistics with global km-scale models remain prohibitively expensive \cite{Don2024}.

Downscaling is a strategy to bridge the gap between affordable coarse resolution climate models and expensive km-scale simulations \cite<e.g.>{Giorgi2019ThirtyYO, LucasPicher2021ConvectionpermittingMW, rampal_enhancing_2024}.  Dynamical downscaling uses the output from a coarse-grid climate model as boundary conditions to drive a separate km-scale regional model. This is also computationally expensive, though less so than global km-scale modeling because the regional model covers only a small fraction (typically only a few percent, \citeA<e.g.>{Katragkou2024DeliveringAI}) of the globe.  Statistical downscaling uses empirical statistical models operating on coarse-grid model outputs to produce fields or statistics of interest (e.g. precipitation or temperature extremes).  It is typically computationally inexpensive but must be customized to the region of interest and may be difficult to adapt to the diverse spatial structures of weather events that can occur there, e.g. tropical cyclones and midlatitude storms in North Carolina, making it potentially less reliable for extreme weather events. Both of these methods rely on the availability of data outputs from a parent climate model.  They can give a representative sample of internal weather variability only if an adequate ensemble of simulations is available for the period of interest.

Machine learning (ML) provides a unique opportunity to leverage km-scale global simulations into an exciting new approach to climate modeling.  Recent ML emulators of the atmosphere with O(100~km) grid spacing are highly computationally efficient and have proven skillful for both weather and climate timescales \cite{Weyn2020, Keisler2022, wattmeyer23, wattmeyer25, cresswellclay2025deeplearningearthmodel}.  Super-resolution using diffusion-based generative ML promises the speed of statistical downscaling and the fidelity of dynamical downscaling \cite{addison2024,mardani_residual_2024}.  Our vision is to train an autoregressive machine learning emulator on sufficiently long samples of training data coarse-grained from simulations with a km-scale reference model.  

In tandem, we would use outputs of the km-scale climate simulations to train an ML downscaling model that skillfully and efficiently maps from  coarse-grid emulator outputs anywhere in the world back to generated samples of a societally important km-scale field such as precipitation.   Both models should produce unbiased probability density functions (PDFs) of their outputs so they can be developed independently then easily used together without further finetuning.  This combination of ML emulator and ML downscaling would allow using the highest fidelity atmospheric models that we have for local climate adaptation planning.  

Recent studies have explored elements of this vision. GenFocal \cite{wan_regional_2025} presents a two-step process applied to existing climate model outputs from the CESM large ensemble. In this case, the model debiases coarse field outputs from CESM ($\sim$150 km) and then performs a super-resolution step to generate 25 km outputs analogous to ERA5 data.  This model accurately recover statistics of tropical cyclone genesis and track locations, and achieves good skill with representing compound (multi-variate) extreme events. CBottle \cite{brenowitz_climate_2025} takes a similar two-step approach, but actually provides an emulator of a coarse resolution climate model (trained on both ERA5 and high resolution ICON data coarsened to 100 km), along with a separate model to downscale those coarse outputs to represent ICON 5 km resolution outputs.  Instead of the typical temporal step-based autoregressive rollout, they use an alternative modeling approach, a fully probabilistic generative diffusion model.  Conditioning on the time of day, day of year, and SST forcing, this model generates direct multivariate samples from the target climate distribution.

The goal of this work is to implement our vision within the framework of a computationally efficient coarse-grid autoregressive seamless weather/climate emulator of a km-scale global atmosphere model.  We present HiRO-ACE, the High Resolution Output Ai2 Climate Emulator.  It is trained on a decade-long simulation of a 3 km global atmosphere model, X-SHiELD \cite{Cheng2022,Har2023} from NOAA-GFDL, forced by historical sea-surface temperatures.  We use a stochastic variant of ACE2 \cite{wattmeyer25}, which we call ACE2S, as an autoregressive climate emulator producing 100 km outputs which can be fed into a diffusion model (HiRO) for generating 3 km precipitation outputs---a 32x downscaling---for any desired time range and region.  

A stochastic emulator is the right choice for this application because it aims to generate a random sample from the PDF of possible weather states conditioned on its state and boundary forcing from the previous rollout step.  That supports our design goal of correctly simulating the grid-scale PDF of the precipitation field.  In contrast, a deterministic emulator averages over those weather states, producing outputs (including precipitation fields) that are somewhat blurred at the grid scale \cite<e.g.,>{lang24,price25}.  Like physics-based climate models, the autoregressive framework provides the full time evolution of weather/climate events of interest. The diffusion model also stochastically generates unbiased random samples from the PDF of the 3 km precipitation field with realistic spatial organization, conditioned on the atmospheric state on the 100 km grid.  Hence, it works together with the stochastic emulator to generate unbiased 3 km precipitation samples consistent with the emulator rollout.  

In Section \ref{sec:hiro-ace-overview}, we provide an overview of our framework and high-level results. In Section \ref{sec:shield-simulation} we discuss the X-SHiELD simulation used for training and evaluation data.  In Sections \ref{sec:ace2s} and \ref{sec:hiro-downscaling}, we detail the climate emulator and downscaling model respectively, followed by comprehensive evaluation of the end-to-end system.

\section{HiRO-ACE Overview}
\label{sec:hiro-ace-overview}

\subsection{HiRO-ACE Framework}
\label{sec:overview-framework}

\begin{figure}[h]
\centering
\includegraphics[width=\textwidth]{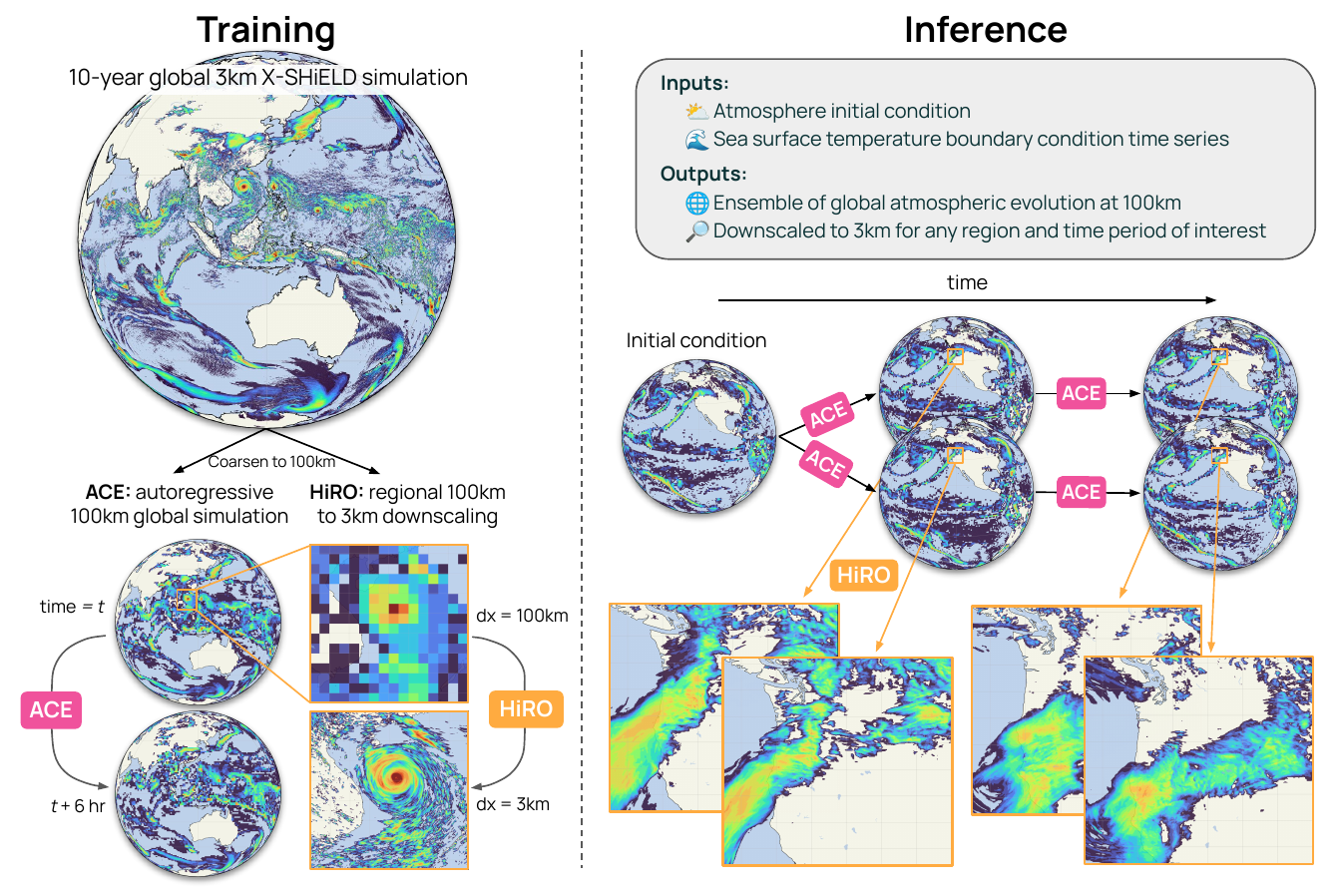}
\caption{Schematic of the two-stage HiRO-ACE framework. Left: using output from a 10-year global 3 km resolution X-SHiELD simulation, we train ACE, an autoregressive global climate emulator, and HiRO (Hi-Resolution Output) a regional downscaling model. Right: these models can be used together to quickly generate large ensembles of precipitation predictions at 3km resolution nearly anywhere on the globe.}
\label{fig:schematic}
\end{figure}

HiRO-ACE combines two machine learning components to emulate 3 km climate simulations: a stochastic 100 km resolution global climate emulator (ACE2S; ``stochastic'' ACE2), and a stochastic downscaling model (HiRO; High Resolution Output) that maps from 100 km inputs to 3 km outputs anywhere on the globe except the polar regions. Figure \ref{fig:schematic}  shows the overall HiRO-ACE framework for training and inference. Both ML models are trained on the same 6-hourly outputs from a reference global km-scale X-SHiELD simulation (Fig.~\ref{fig:schematic} left).  To minimize data storage requirements, an important consideration for km-scale simulations, we coarsen three-dimensional X-SHiELD outputs in-line to an intermediate coarser grid, and preserve selected native resolution X-SHiELD surface fields (e.g. precipitation) to train the downscaling model. For prediction (Fig.~\ref{fig:schematic} right), we leverage the computational efficiency of ACE2S, to run autoregressive climate simulations for a period of interest, and then downscale selected regions with HiRO.   Both components are probabilistic, allowing a user to generate random realizations of the large-scale weather/climate variability (ACE2S) and the small-scale variability for individual snapshots (HiRO). We note that training each machine learning component separately was sufficient and no specific fine-tuning of HiRO using ACE2S output was necessary.

\subsubsection{Model components}
ACE2S is trained in a two stage process to generate 6-hourly autoregressive predictions of atmospheric fields. The pretraining stage uses a longer historical reanalysis dataset, ERA5 \cite{Her2020}, to establish the foundation of the model’s dynamics. Then the model is fine-tuned on coarsened data from the 3 km X-SHiELD simulation.  The major modification from the previous version of ACE is the added ability to condition the model on injectable noise through a conditional layer normalization that allows training on a probabilistic loss function (see Section \ref{sec:ace2s-model-arch}). These changes help ACE2S preserve the appropriate spatial variability characteristics of precipitation for use with the downscaling model, while maintaining the correct climate.

The downscaling model (HiRO) is trained on 100 km surface wind and precipitation inputs to produce 3 km precipitation fields consistent with the original physical model, X-SHiELD. We use the CorrDiff \cite{mardani_residual_2024} approach where we train a diffusion model to make a correction to a mean field estimate.  While the original CorrDiff methodology uses another ML model to predict the initial mean, we instead use a simple bicubic interpolation of the coarse input fields.  This approach works well based on a variety of “perfect prediction” tests (see Section \ref{sec:hiro-downscaling} for model and performance details).

\subsubsection{Computational requirements}
The framework is computationally efficient to run. The inference speed of ACE2S is comparable to the original deterministic ACE2 model (1500 simulated years per day for $1^\circ \times 1^\circ$ grid resolution on one NVIDIA H100 GPU). 

Downscaling a single 16$^{\circ}\times$16$^\circ$ patch of one year of ACE2S output takes approximately 45 minutes on an H100 GPU; this could likely be reduced by an order of magnitude by more careful optimization of the downscaling model (e.g., through distillation of the denoiser \cite<e.g.,>{salimans2022progressivedistillationfastsampling}).

The 10-year km-scale reference simulation is by far the most expensive and time-consuming part of training the ML models, requiring more than 800 GJ of energy  (see Section 3), but it also has many other scientific applications. Training ACE2S and HiRO takes 7 days and 4 days, respectively, on 8 H100 GPUs.  The combined energy cost of training both models is around 1\% ($\sim$6--8 GJ) the cost of the X-SHiELD simulation, while the cost of inference to produce a 10-year downscaled global 3 km dataset is around 0.5\% ($\sim$4 GJ) of the X-SHiELD simulation.  However, most common use cases would not require global downscaling, so real inference costs are likely much lower.

\subsection{Key Results}
\label{sec:overview-key-results}

\begin{figure}
\centering
\includegraphics[width=\textwidth]{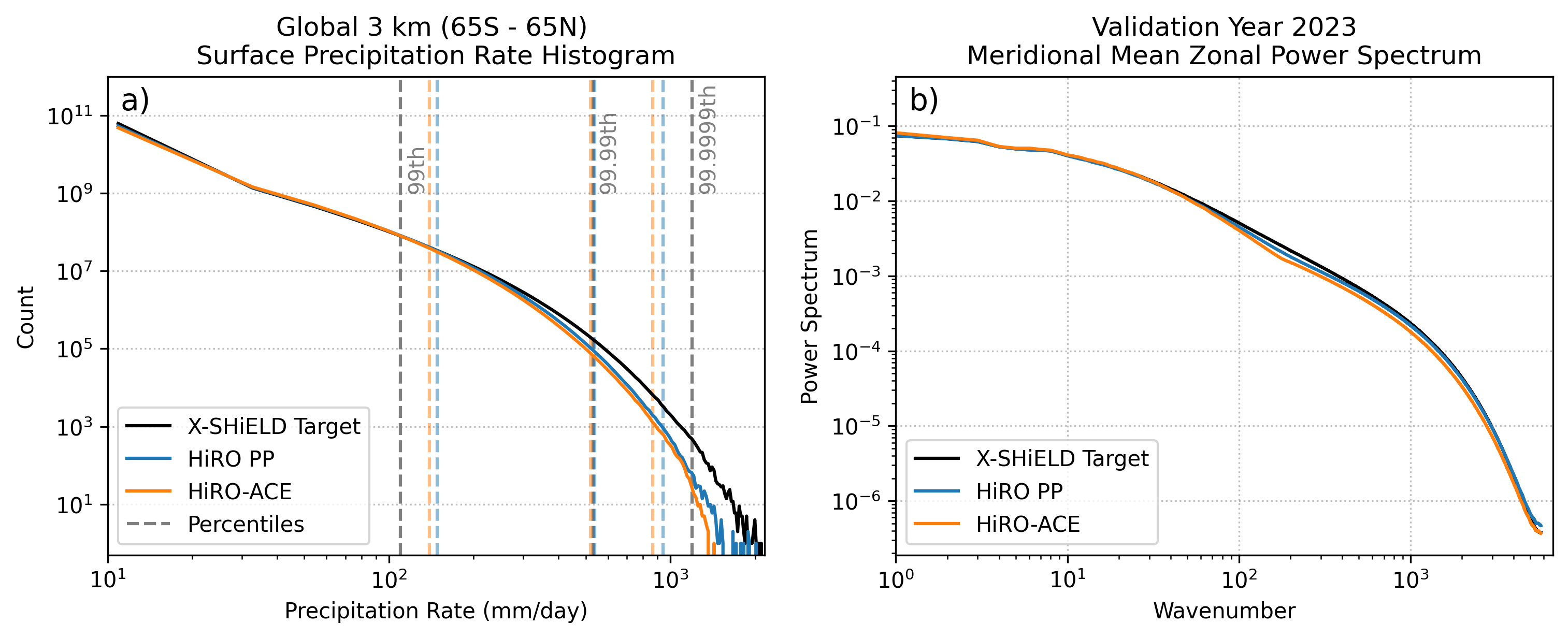}
\caption{Global (\SI{65}{\degree S}--\SI{65}{\degree N}) fine-grid 6-hourly precipitation statistics during the 2023 hold-out year for X-SHiELD, a single HiRO ``perfect prediction'' ensemble member, and a single HiRO-ACE ensemble member: (a)  Histograms of binned 6-hr average surface precipitation rates over all 3 km grid columns and sampling times, with the 99th, 99.99th, 99.9999th percentiles of the data marked with vertical dashed lines;  (b) time and meridional mean of the zonal power spectrum.}
\label{fig:histogram-ace-hiro}
\end{figure}

\subsubsection{Accurate frequency of km-scale extreme precipitation}
First, to be useful for assessing climate impacts, HiRO-ACE should accurately match the PDF of km-scale precipitation simulated by X-SHiELD, especially in the extremes. Figure \ref{fig:histogram-ace-hiro} shows a precipitation rate histogram comparison over the 1-year validation period for the near-global (\SI{65}{\degree S}--\SI{65}{\degree N}) domain. In this figure, we compare results from a 1-year free-running ACE2S simulation (over 2023) with precipitation output downscaled using HiRO (HiRO-ACE), “perfect prediction” (PP) where HiRO downscales directly from 100 km coarsened X-SHiELD data, and the reference X-SHiELD 3 km histogram. The perfect prediction results show the histogram errors that are due purely to HiRO, and any additional bias in HiRO-ACE can be attributed to ACE2S. 

In Figure \ref{fig:histogram-ace-hiro}, both the HiRO PP and HiRO-ACE 3 km precipitation histograms display close correspondence to each other into the far tails of the precipitation distribution for the near-global domain.  This highlights that ACE2S successfully maintains the output distribution expected by HiRO, which is trained only on coarsened X-SHiELD data. If the drift were too large or there were large changes in the character (e.g., power spectrum, discussed in Section \ref{sec:ace2s-results}), the downscaling output fidelity would collapse. Compared to the target X-SHiELD histogram, HiRO PP and HiRO-ACE show good correspondence to the target out into the 99.99th percentile ($\sim$500 mm/day) of the distribution.  For the most extreme X-SHiELD rainfall ($>$1000 mm/day), HiRO PP and HiRO-ACE global downscaled outputs display fewer generated instances with the calculated 99.9999th percentiles only reaching 78\% and 73\% of the X-SHiELD target, respectively.  Many of the most extreme precipitation events are isolated convective storms over tropical oceans (SI Fig. S1), which appear as muted precipitation rates ($\sim$100--200 mm/day) when coarsened to 100 km. At coarse resolution, the moderate precipitation rates in the tropics rarely correspond to extreme 3 km precipitation events (only about 0.2\% of the total events).  The relative rarity in the data is a likely factor for the downscaling model's underestimation in the far tail of the distribution. 

\begin{figure}
\centering
\includegraphics[width=\textwidth]{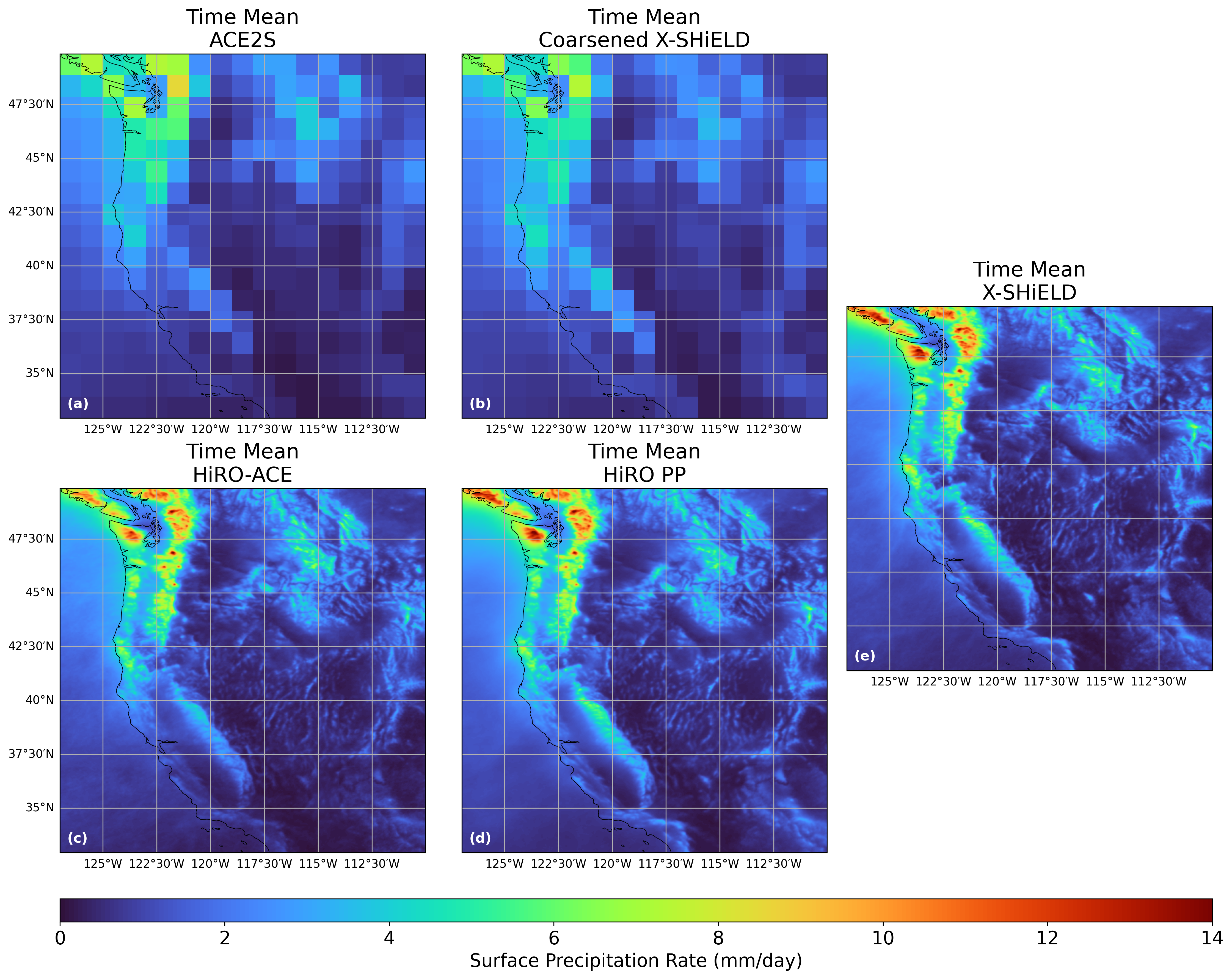}
\caption{Annual-mean precipitation over the 2023 hold-out year for (a) a single selected ensemble member of ACE2S, (b) coarsened X-SHiELD, (c) one HiRO-ACE ensemble member, (d) one HiRO perfect prediction ensemble member, and (e) fine-grid X-SHiELD.}
\label{fig:west_coast_time_mean}
\end{figure}

\subsubsection{Accurate km-scale spatial pattern of time-mean precipitation}
\label{sec:hiro-ace-time-mean}
Second, HiRO-ACE should be able to replicate the km-scale spatial pattern of time-mean precipitation from X-SHiELD, even in regions of complex geography, ideally to within fluctuations due to internal weather variability. Figures \ref{fig:west_coast_time_mean}a-b compare the spatial pattern of annual-mean precipitation over the western U.S. for a single 1-year ACE2S simulation against coarsened X-SHiELD data over the 2023 hold-out year. The ACE2S free-running simulation captures the overall character of the annual-mean precipitation features seen in coarsened X-SHiELD. However, there is a slight dry bias in the southwestern U.S. compared to coarsened X-SHiELD.  The natural internal variability of precipitation averaged over this one-year hold-out period is too large to test if this bias is meaningful. A longer ACE2S simulation discussed in Section \ref{sec:ace2s-results} suggests that it is partly due to model bias.  

Figures \ref{fig:west_coast_time_mean}c-e compare the corresponding downscaled HiRO-ACE, HiRO PP, and target X-SHiELD fields.  The HiRO PP time mean is virtually indistinguishable from the X-SHiELD target, showing most of the bias is a result of the 1-year ACE2S differences from X-SHiELD for this region.  The dry bias from ACE2S in the southwestern U.S. leads to a decrease in the time-mean precipitation on the windward side of this region's orography in the time-mean HiRO-ACE map. 

Although both HiRO and ACE are trained globally, the detailed spatial pattern of the downscaled 3 km annual-mean precipitation from both models is remarkably accurate over the complex topography of this region.  Global time-mean and bias maps (SI Fig. S2 and S3) for HiRO-ACE and HiRO PP over the 2023 hold out period show characteristics similar to this regional analysis, that is, small global mean biases with larger HiRO-ACE pattern errors due to the ACE2S and X-SHiELD mismatch.  We further discuss the holdout 2023 and 10-year biases of HiRO-ACE and HiRO PP in Section \ref{sec:hiro-ace-time-mean}.  

\subsubsection{Well-reconstructed km-scale precipitation features}
Last, we look at the ability of HiRO to downscale features that are poorly resolved at the 100 km grid scale simulated by ACE2S.  We consider a 6-hour ACE2S forecast of a representative tropical cyclone simulated by X-SHiELD near southern China.  Because the forecast is so short, the HiRO downscaling of this forecast can meaningfully be compared with the X-SHiELD ``truth''.  Even with a perfect prediction of the coarse 100 km atmospheric state, it is impossible to deduce the exact structure of the underlying cyclone, such as the eye and the km-scale detail of the organized precipitation features around it.  Grid-scale errors or stochastic noise in the 6-hour ACE2S rollout magnify this challenge. However, with 10 years of X-SHiELD training data including over 500 tropical cyclone examples from around the world, HiRO draws from this distribution to generate a plausible cyclone event. ACE2S generates coarsely-resolved tropical cyclone-like disturbances throughout multi-year rollouts, so these downscaled results are broadly representative of free-running simulations using our framework. 

\begin{figure}
\centering
\includegraphics[width=\textwidth]{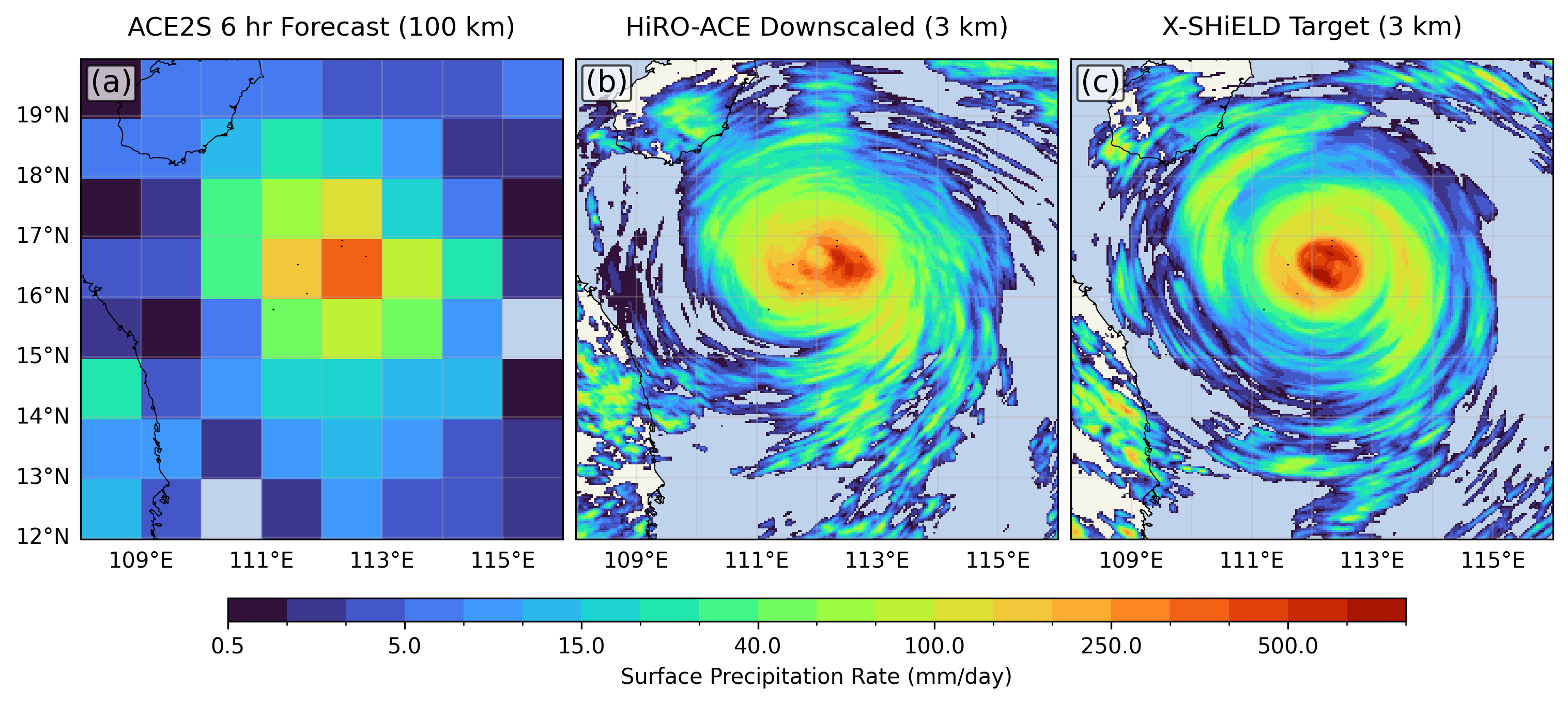}
\caption{Tropical cyclone case study near southern China showing (a) ACE2S 6-hour forecast of 6 h average surface precipitation rate at 100 km resolution, (b) HiRO downscaled output at 3 km resolution, and (c) target X-SHiELD precipitation at 3 km resolution.}
\label{fig:hurricane-focus-ace-hiro}
\end{figure}

Figure \ref{fig:hurricane-focus-ace-hiro}a shows one realization of an ACE2S 6-hour forecast of the average surface precipitation rate over the South China Sea initialized from the coarsened X-SHiELD atmospheric state at the previous timestep. At 100 km resolution, the cyclone appears only as a radially decreasing blob of precipitation with a central maximum rainfall rate in the vicinity of 300 mm/day.  Using ACE2S output, HiRO generates features (Fig. \ref{fig:hurricane-focus-ace-hiro}b) consistent with the target X-SHiELD cyclone (Fig. \ref{fig:hurricane-focus-ace-hiro}c), showing a cohesive inner core with maximum rainfall $>$500 mm/day and separated outer rainband features. The downscaled sample does not show the outer edge of the eyewall as clearly, but we do not expect the generated output to exactly match the target, especially for rare events such as tropical cyclones, which can have a range of intensities and morphological details. See SI Figure S4 for examples of other realizations of HiRO-ACE outputs for this case.

\section{11-year X-SHiELD reference simulation}
\label{sec:shield-simulation}

The primary reference data for emulation and downscaling is derived from an 11-year physics-based simulation using the global-storm-resolving X-SHiELD model developed at the Geophysical Fluid Dynamics Laboratory (GFDL).  This is one of the longest simulations to date with a global storm-resolving model (GSRM). % with a horizontal grid spacing of less than 5~km. 
GSRMs differ from traditional coarse-resolution climate models in that due to their fine horizontal resolution (less than \SI{5}{\km}) they do not require a parameterization for the effects of deep convection, often cited as a source of uncertainty in climate simulation, and explicitly resolve the effects of fine-scale topography \cite{Ste2019,Tak2024,Seg2025}.   

X-SHiELD uses a \SI{~3.25}{\km} resolution cubed-sphere grid in the horizontal and 79 hybrid-sigma-pressure layers in the vertical \cite{Cheng2022,Har2023}.  The simulation was initialized from GFS analysis on 00Z 2013-01-01, and was forced by sea surface temperatures and sea ice concentration from ERA5 reanalysis \cite{Her2020} and observed carbon dioxide concentration from the NOAA Global Monitoring Laboratory \cite{Con1994}. The dynamical core and physical parameterization settings used were similar to those of \citeA{Cheng2022}, with the exception of an upgrade to version 3 of the GFDL Microphysics \cite{Zho2022}, and further tuning to achieve improved top of atmosphere and surface radiative fluxes, marine stratocumulus cloud coverage, and tropical cyclone activity.

The simulation was run on the C6 partition of GFDL's Gaea computer, using \num{27648} cores on AMD EPYC 9654 processors with a throughput of \num{0.12} simulated years per day, corresponding to an energy usage of \SI{74.6}{\giga \joule} per simulated year. The overall energy cost of this simulation exceeds \SI{800}{\giga \joule}, which dwarfs the approximately \SI{10}{\giga \joule} required for training and running our ML models for a similar output length.  To put these energy costs in perspective with those of other large-scale computing projects, training GPT-3, in its time a state-of-the-art large language model, used roughly \SI{4600}{\giga \joule} \cite{Pat2021}.

For the purposes of this work, we treat the first year of the simulation as spin-up and only use data from the final 10 years for training, testing, and statistics calculations. We used the first 9 years to train both the downscaling and coarse, global model, leaving the final year (2023) for independent testing of the full pipeline. Due to internal weather variability, a one-year testing period is quite short for some purposes, such as comparing ML-generated time-mean precipitation with the reference km-scale model.  For certain cases we compared 10-year rollouts of the ML framework with the 10 retained years of X-SHiELD data.  Because our ACE2S emulator is trained to optimize very short-term forecasts, even comparing it with a sufficiently long in-sample period of X-SHiELD is a strong test of the quality of its precipitation climatology.

Training on only 9 years of X-SHiELD data provides a limited number of reference samples for autoregressive emulation and requires pretraining on a separate dataset for best results. Nevertheless, the unique dataset produced by this simulation presents the opportunity to train machine learning models for emulation and diffusion-based downscaling in a fully consistent way.

Emulation-related variables, the same 6-hourly fields as those required for ACE2 in \citeA{wattmeyer25}, were derived from outputs that were horizontally coarsened online to C384 (\SI{~25}{\km}) resolution using area-weighted block averaging.  For compatibility with ACE2S, these fields were regridded offline to a \SI{1}{\degree} Gaussian grid using the first-order conservative remapping scheme of GFDL's \texttt{fregrid} tool \cite{NOA2024} and vertically coarsened using mass-weighted block averaging to ACE2S's 8 finite-volume vertical layers (see the ``SHiELD'' column in Table S2 of \citeA{wattmeyer25} for the exact vertical layers used). Prior to vertical coarse-graining, ACE2S prognostic variables were run through a spherical harmonic transform round trip to mitigate the impact of sharp gradients introduced by the conservative remapping scheme at high latitudes.

Fine-resolution downscaling-related variables (the 6-hourly interval-average total precipitation rate and static surface height) were derived from outputs at the native C3072 (\SI{~3.25}{\km}) resolution.  For exact alignment with the edges of ACE2S's \SI{1}{\degree} Gaussian grid, these fields were conservatively remapped using \texttt{fregrid} to a custom 5760 x 11520 grid constructed by uniformly refining the 180 x 360 Gaussian grid by a factor of 32 within each cell.

\section{Stochastic ACE2S trained on X-SHiELD}
\label{sec:ace2s}

\subsection{Model architecture and design}
\label{sec:ace2s-model-arch}

Our downscaling strategy assumes that our coarse-grid climate emulator will provide the ML downscaling model with fields of precipitation (and other downscaling inputs) that have realistic grid-scale spatial variability that is consistent with coarsened output from our km-scale reference model.  Like deterministic ML weather forecast models, ACE2 is overly diffusive at the grid scale (Fig. ~\ref{fig:power_spectrum_comparison}). Some stochastic weather AI models such as AIFS-CRPS \cite{lang24} have shown promise in addressing this issue, which is critical for our application, without greatly slowing down inference.

To this end, we developed ACE2S. This is a modified version of ACE2 \cite{wattmeyer25}, a deterministic model which uses a Spherical Fourier Neural Operator architecture \cite{bonev23} including architectural constraints to ensure conservation of dry air mass and moisture. Following GenCast \cite{price25} and FGN \cite{alet25}, we introduce stochasticity by using conditional layer normalization instead of the instance normalization used by ACE2. Layer normalization normalizes features across input channels independently for each column, instead of independently across the horizontal domain as in \citeA{wattmeyer23} and \citeA{wattmeyer25}. Model predictions are conditioned on 64 isotropic Gaussian white noise channels, as in \citeA{price25}.

Several weather AI models have implemented stochastic prediction using cumulative ranked probability scale (CRPS, \ref{eq:crps-definition}) in their loss function. Models that use CRPS without filtering strategies or additional loss terms (e.g., FGN \cite{alet25} and Swift \cite{swift}) report excess spectral power at small scales. AIFS-CRPS \cite{lang24} ameliorates this noise by filtering small scales in the residual component of the architecture, while FourCastNet 3 \cite{bonev25} adds the CRPS of the absolute spectral power of the spectral coefficients as an additional loss term. The stochastic version of NeuralGCM \cite{kochkov24} includes both CRPS and the energy score of complex spectral coefficients of spherical harmonic components of predicted fields (which they call spectral CRPS).  This spectral energy score treats complex spectral components as values in $\mathbb{R}^2$ and is a ``fair" score. NeuralGCM includes this second term only for wavenumbers below 80 and does not report the spectral power behavior of the stochastic model.

Our loss function builds on these prior works using a combined nodal CRPS and spectral energy score similar to that used in NeuralGCM \cite{kochkov24}, but without a maximum cut-off wavenumber, without area weighting for the nodal component, and using ``almost fair" CRPS for the nodal component \cite{lang24} with $\alpha=0.95$. We scale the energy score by an empirical factor of $(2 n_l)^{-1}$, where $n_l = 180$ is the number of l-values in the spectral space, to grant similar magnitude to the CRPS when evaluated on Gaussian-distributed random variables, regardless of domain size. Following this, we empirically choose weights of $0.9$ on the nodal CRPS and $0.1$ on the spectral energy score. Losses are computed on variables normalized as in \citeA{wattmeyer25}, including per-variable weights. While the ACE2S model shares roots with FourCastNet, especially the use of SFNO blocks \cite{bonev23}, the conditional SFNO and changes made to training methodology here are distinct from FourCastNet 3 \cite{bonev25}.

\subsection{Training Strategy}

GFDL's 10-year X-SHiELD reference simulation is a computational tour de force, but when used alone it is barely sufficient to train ACE2S. We found that we could significantly reduce ACE2S's climate biases by pretraining it on a longer but comparable dataset, using years 1979-2022 from the ERA5 \cite{Her2020} dataset described in \citeA{wattmeyer25}. During this 120-epoch stage, we use a one-step prediction loss. We use the same training and validation methodology as ACE2-ERA5, except that the checkpoint with the lowest validation loss is selected instead of the one that has the best climatology during the testing period. 

From this starting point, we train ACE2S on the X-SHiELD data using a multi-step loss. We train for 120 epochs using an ensemble size of 2, a constant learning rate of $10^{-4}$, an AdamW optimizer with weight decay of 0.01, and EMA with weight decay of 0.999. Because we only use 10 years of X-SHiELD data vs. the 44 years of ERA5, but train both stages for the same number of epochs, the multi-step training stage involves about four times fewer gradient descent calculations than the single-step pretraining.

Similar to several weather forecasting ML models \cite <e.g.,>[]{lang24,bonev25}, we augment the training data using autoregressive model output by randomly selecting 1 step (6 hours), 2 steps, 4 steps, 12 steps, or 20 steps (5 days) of rollout for each batch, with probabilities of 60\%, 20\%, 10\%, 5\%, and 5\% respectively. This empirical choice balances training efficiency vs. the potentially improved climate fidelity of multi-step rollouts, while not degrading the 1-step predictions. We auto-regress the model from two identical initial conditions over the chosen rollout period, but back-propagate and optimize the model prediction over only the final 6-hour timestep (by which time the two states have diverged). This ``pushforward trick'' first introduced in \cite{brandstetter2023messagepassingneuralpde} is more computationally efficient than back-propagating through the entire window and is similar to the training strategy used in the Aurora model \cite{Bodnar2025-Aurora}.

As with ACE2, during the X-SHiELD multistep training stage the model was evaluated after every epoch using a 4 member ensemble of identically initialized inference runs for the first 5 years of the training data (2014 - 2018). For robustness, we trained 4 different random seeds with this training strategy.  For each random seed, we saved the best checkpoint based on the channel-mean of global root mean squared time-mean bias (RMSB) calculated using these 5-year runs. We found that 3 of the 4 random seeds performed similarly, with 1 random seed having significantly higher biases with respect to X-SHiELD. For the 3 corresponding saved checkpoints, a 10-year run (9 years of training data and 1 year of out-of-sample) was compared to X-SHiELD. Of these three checkpoints, the best was chosen to jointly optimize channel-mean error in the time mean bias and the bias in the spectral power of precipitation at the Nyquist wavenumber of the spherical harmonic transform used by SFNO (i.e. with half-wavelength equal to the $1^\circ$ meridional grid spacing).

Due to the need for two ensemble members, ACE2S training requires twice as much data processing per training example as for ACE2. This is mitigated by using a single forward step during the initial phase, and a batch size of 8 instead of 16 (the loss is reduced nearly as fast with the smaller batch size). Taking advantage of speedup due to only training on the final step and data-loading optimizations, the multi-step training is only 3 times slower per batch than the 1-step training. The inference speed of ACE2S is comparable to the original deterministic ACE2 model (1500 simulated years per day for $1^\circ \times 1^\circ$ grid resolution on one H100 GPU).  For more details on the sensitivity of the ACE2S training choices, see SI Text S1 and S2.

\subsection{ACE2S improves on ACE2}
\label{sec:ace2s-results}

In this section, we show that when trained on coarsened X-SHiELD data as described above, ACE2S stably reproduces its time-mean climatology and variability with minimal biases, maintains the desired grid-scale variability of precipitation, and improves on our deterministic ACE2 model in both these regards.

For this comparison, we similarly train ACE2. Specifically, we start with the best model checkpoint from ACE2-ERA5 and train for 120 epochs on the coarsened X-SHiELD dataset with the same two-step mean-squared error loss and variable weightings as in \citeA{wattmeyer25}.
Due to the limited amount of X-SHiELD data (10 years), only the last year was set aside for independent testing.  It is difficult to determine climate skill with just 1 year of test data due to internal variability. Hence, we evaluate both models on a 10-year run (9 years of training data plus the 1 year of test data) starting on 00Z January 1, 2014. ACE2S reproduces the climate of coarsened X-SHiELD remarkably well, with RMSBs 30-80\% smaller than ACE2 for all predicted variables (see SI Fig.~S5). For the variables used by HiRO (zonal and meridional components of 10 meter wind and the surface precipitation rate) the global area-weighted RMSBs for the 10-year period for ACE2S (ACE2) are  0.18 (0.32) m/s, 0.13 (0.20) m/s, and 0.23 (0.48) mm/day. 

\begin{figure}
\centering
\includegraphics[width=\textwidth]{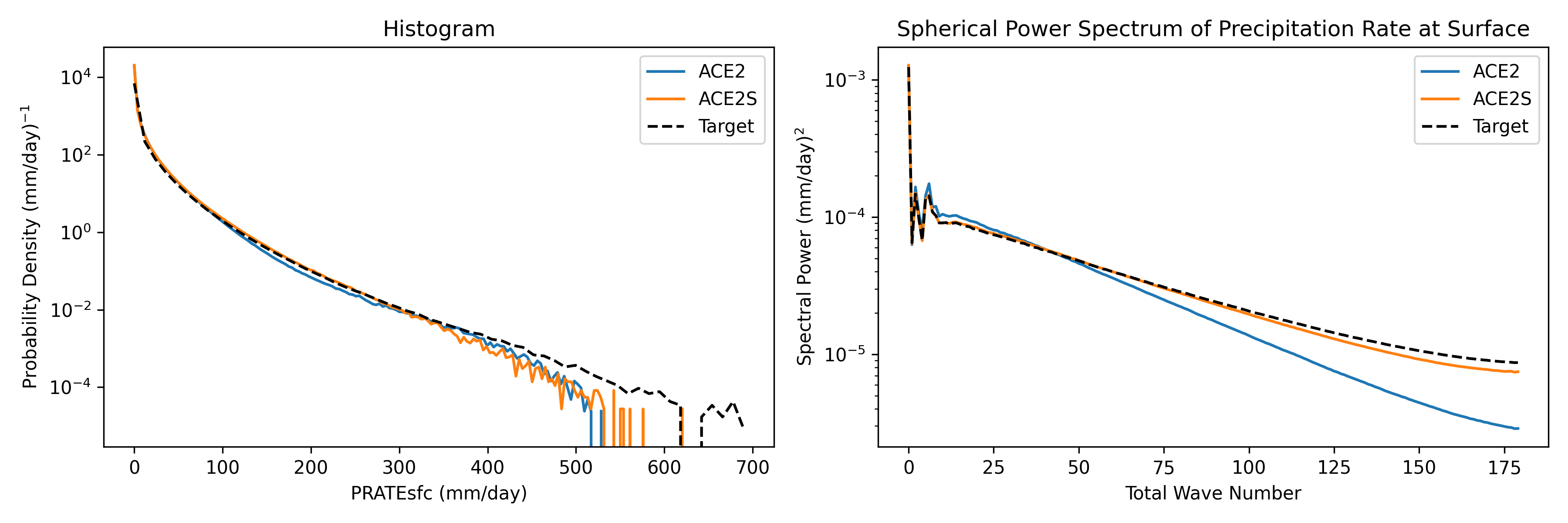}
\caption{Comparison of surface precipitation (a) histograms and (b) spherical power spectrum, averaged over all 6-hourly samples of 10-year runs of deterministic ACE2 (blue), ACE2S (orange), and the coarsened X-SHiELD target.}
\label{fig:power_spectrum_comparison}
\end{figure}

The coarse-grid simulation needs to maintain the correct grid-scale precipitation spatial variability and intensity to be consistent with HiRO's ``perfect prediction'' training protocol. ACE2 underpredicts spectral power by a factor of three at the smallest scales resolved by its grid.  We attribute this to smoothing encouraged by its MSE loss and 2-step training.  ACE2S, which uses a probabilistic loss function, nearly removes this bias (Fig. ~\ref{fig:power_spectrum_comparison}a) and  modestly improves the most extreme precipitation rates (Fig. ~\ref{fig:power_spectrum_comparison}b). 

\begin{figure}
\centering
\includegraphics[width=\textwidth]{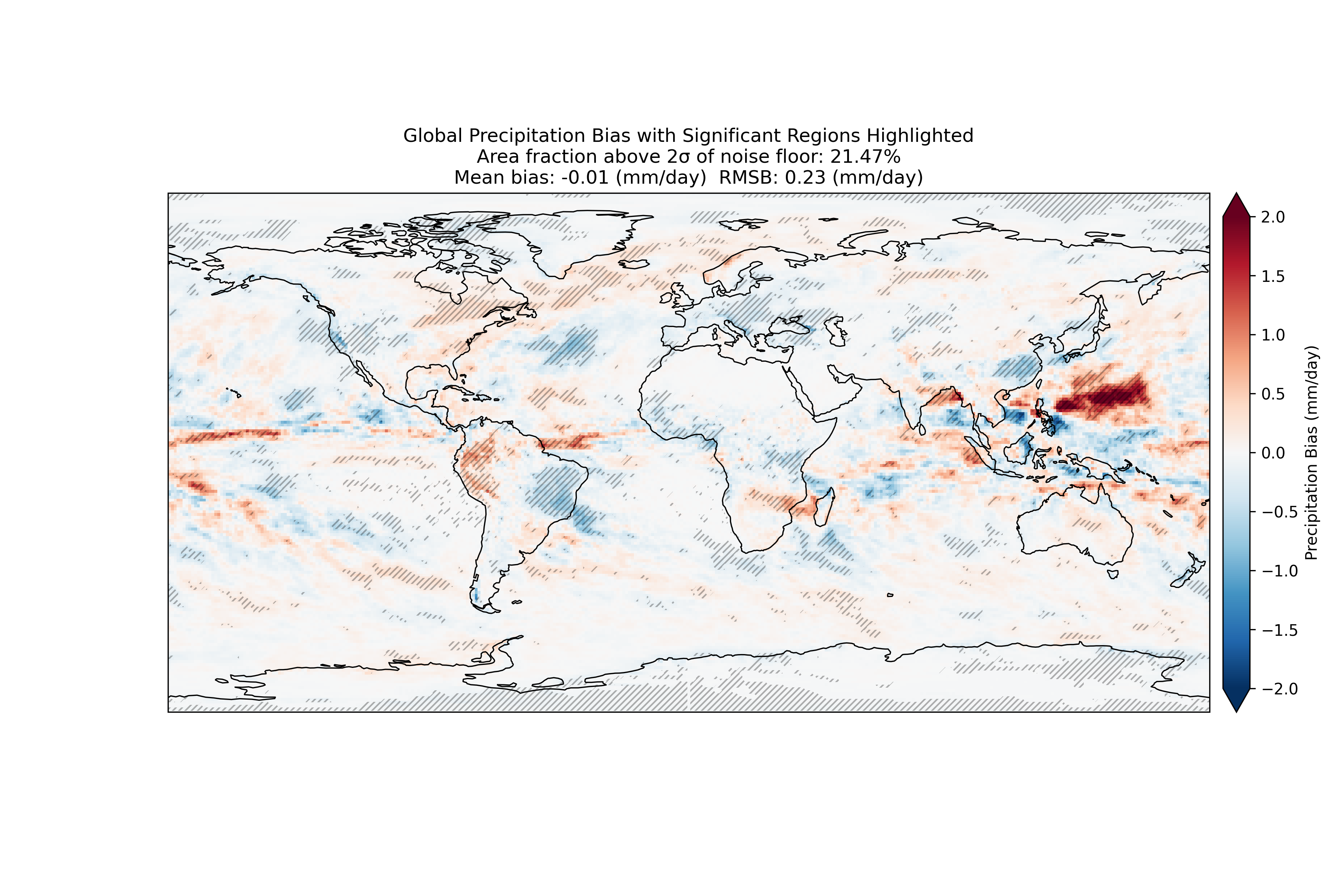}
\caption{Global time-mean bias plot of the 10-year ACE2S run compared to 10 years of coarsened X-SHiELD (2014-2023). Hatching indicate gridpoints where ACE2S absolute biases are at least 0.1 mm/d and exceed 2 standard deviations of the noise floor due to unforced internal variability. }
\label{fig:global_precip_bias}
\end{figure}

Figure \ref{fig:global_precip_bias} shows the global 10-year mean surface precipitation biases for the ACE2S simulation, hatched in regions (collectively covering about one quarter of the globe) where they are larger than 0.1 mm/day and a ``noise floor'' analysis (see SI Text S3 for details) suggests they exceed (at 95\% confidence) the likely range of natural internal variability. 

Most of the biases are small, on par or lower than other versions of ACE (e.g., ACE-ERA5), even in hatched regions. There is a notable wet bias over the tropical northwest Pacific. A more in-depth analysis of tropical cyclones (TCs) suggests this bias is associated with overestimated TC frequency (see SI Text S4, Fig. S5 and S6 for more details). Another notable area where ACE2S  significantly underestimates precipitation is the southwestern United States. One potential explanation of these broad biases could be related to small inconsistencies between the ERA5 data used to pre-train ACE2S and the 10-year X-SHiELD dataset used for fine tuning. Further investigation of pre-training and fine-tuning strategies will be investigated in future versions of ACE2S. 

\section{HiRO downscaling trained on X-SHiELD}
\label{sec:hiro-downscaling}

The downscaling model (HiRO) uses a slight modification of the CorrDiff framework \cite{mardani_residual_2024} to train a diffusion-based model for downscaling of ACE2S 100 km outputs to 3 km. Overall, CorrDiff defines a residual prediction strategy for learning the conditional probability density $p(\boldsymbol{x}|\boldsymbol{y})$ of high-resolution output $\boldsymbol{x}$ given input $\boldsymbol{y}$.  This approach simplifies the learning task by decomposing the prediction into two components, $\boldsymbol{x} = \boldsymbol{\mu} + \boldsymbol{r}$. In this decomposition, one model predicts the deterministic mean ($\boldsymbol{\mu}=\mathbb{E}[\boldsymbol{x}|\boldsymbol{y}]$), and another predicts the residual stochastic component, $\boldsymbol{r}$. 

HiRO simplifies the CorrDiff method by using bicubic interpolation
instead of a U-Net regression model to predict the base output component $\boldsymbol{\mu}$.  This makes the diffusion model responsible for predicting  deterministic high-resolution features like the time-mean precipitation interactions with topography along with stochastic high-resolution components of the output fields.  In cBottle, a diffusion-based downscaling model is trained without a residual decomposition to directly predict multivariate high-resolution fields.  We attempted training such a model but were unable to achieve skill comparable to our residual approach with such a model.  This may be due to our larger downscaling factor of 32x compared to past models (e.g., ~6x for GenFocal \cite{wan_regional_2025}, 12.5x for CorrDiff \cite{mardani_residual_2024}, and 16x for cBottle \cite{brenowitz_climate_2025}).  Our interpolation strategy is fast, easy to implement, and only requires training a single downscaling model.

HiRO also differs from CorrDiff and cBottle by using a static high-resolution topography field as a conditioning input rather than including a positional embedding in the mean prediction model.  This approach produces skillful models without any extensive parameter tuning (not shown).  For a multivariate downscaling model with other outputs that might depend on different static model information (e.g., surface air temperature over land might be sensitive to local surface type), it might be advantageous to switch to positional embeddings.

\subsection{Model Training and Inference}

HiRO is trained to predict 3 km surface precipitation rates conditioned on coarse-grid (100 km) inputs of 6 hour average surface precipitation rates and eastward/northward wind at 10 meters.  We chose this minimal set of coarse-grid conditioning features for simplicity; more such features might incrementally improve HiRO accuracy.  To avoid potential issues due to changes in grid spacing at high latitudes, we limited the training inputs to the latitudes 66S--70N.  We train the diffusion denoiser for HiRO using the score-matching approach \cite<e.g., as in>{Song2020ScoreBasedGM, mardani_residual_2024} on the same 9-year segment of X-SHiELD data (2014-2022) used for stochastic ACE.

We train HiRO using a batch size of 24, the Adam optimizer, and a learning rate of 10$^{-4}$.  We train for approximately 740,000 gradient descent steps and use EMA weights (with decay of 0.999) to validate and save checkpoints while training. Like cBottle \cite{brenowitz_climate_2025}, we divide the global paired 100 km and 3 km snapshots into coincident patches to limit GPU memory usage and target reasonable model inference speeds for regional evaluation; we use patches of 16 x 16 (100~km)$^2$ grid cells and 512 x 512 (3~km)$^2$ cells, respectively. (See details of the model parameters, final checkpoint selection, and patching approach for training and inference in \ref{sec:hiro-supplement}).

\subsection{Perfect Prediction Results}

We validate HiRO in a “perfect prediction” (PP) setup, in which we stochastically downscale patches of $16\times16$ (100~km)$^2$ 6-hour-mean precipitation derived from unseen coarsened X-SHiELD data and compare with their paired 3 km targets. These results provide an upper bound on the downscaling skill of our HiRO-ACE framework, which is also affected by the skill of stochastic ACE in simulating the coarse-grid atmospheric inputs to HiRO. 

\subsubsection{Atmospheric river case}
\begin{figure}
    \centering
    \includegraphics[width=\textwidth]{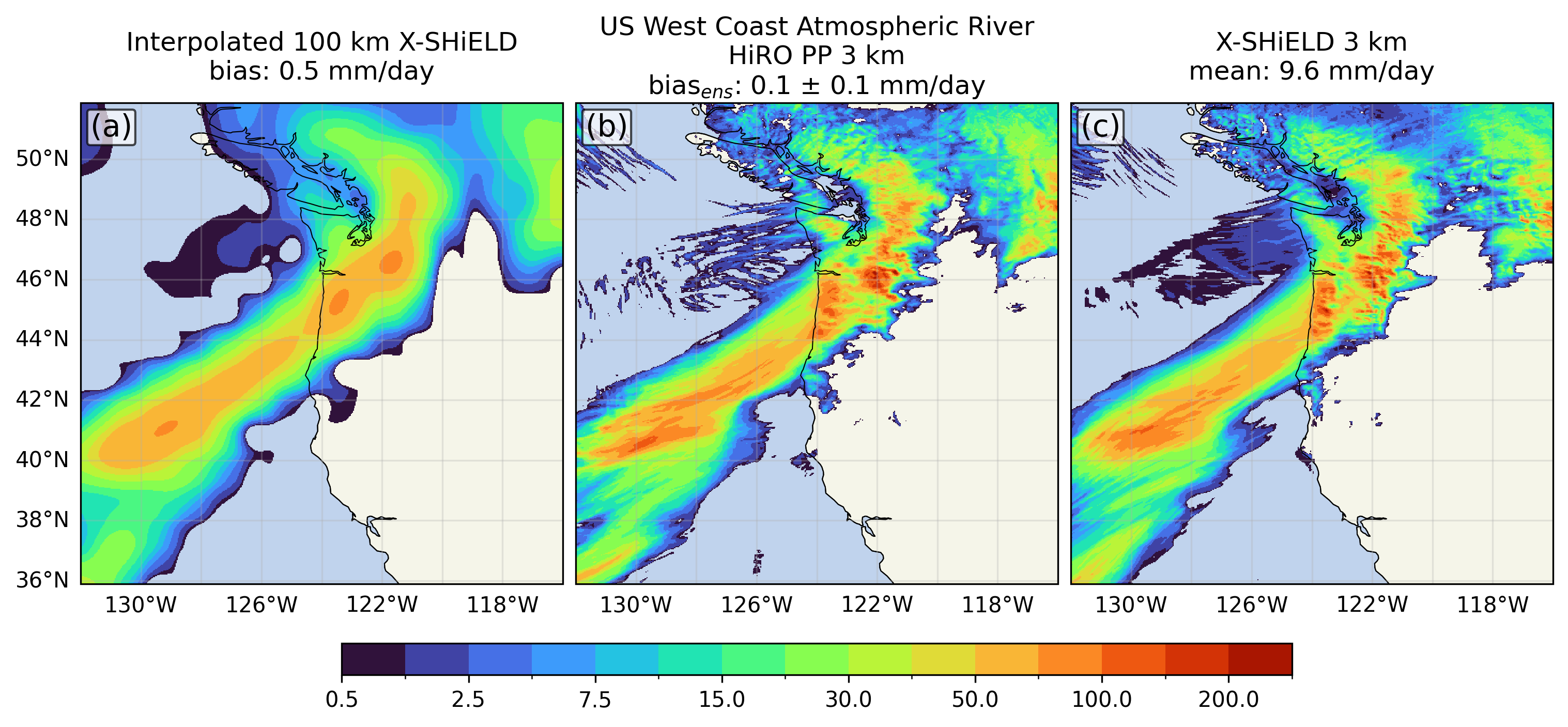}
    \caption{Perfect prediction downscaling example of 6 hr surface precipitation rates for a landfalling atmospheric river along the west coast of the United States showing (a) input bicubic interpolation of the coarse field, (b) HiRO generated sample, and (c) target X-SHiELD 3 km data. Panel titles show (a) the regional bias, (b) the 128-member HiRO PP ensemble mean bias and 95\% confidence interval ($\pm 2\sigma$), and (c) the regional mean.}
    \label{fig:wa-ar-pp}
\end{figure}

Figure \ref{fig:wa-ar-pp} shows a PP downscaling example of an X-SHiELD-simulated landfalling atmospheric river along the west coast of the United States.  Here, warm, moist marine air blowing from the southwest ahead of a cold front precipitates heavily as it encounters the coastal mountain ranges of the western U.S.  The bicubic interpolation (Fig. \ref{fig:wa-ar-pp}a) is the smooth deterministic component derived from the coarse grid precipitation, to which a stochastic correction generated from the diffusion model is added to obtain a downscaled sample (Fig. \ref{fig:wa-ar-pp}b).  This accurately recovers the orographic enhancements and the texture of precipitation over the ocean seen within the front and in the cool showery post-frontal regime in the paired X-SHiELD target (Fig. \ref{fig:wa-ar-pp}c).  Other stochastic realizations of the downscaled precipitation are random draws from the total distribution of possible frontal zones given the coarse-grid conditioning input. They are qualitatively very similar, but not identical to the example shown.

\subsubsection{Tropical cyclone case; multiple stochastic realizations}
\begin{figure}
     \centering
     \includegraphics[width=\textwidth]{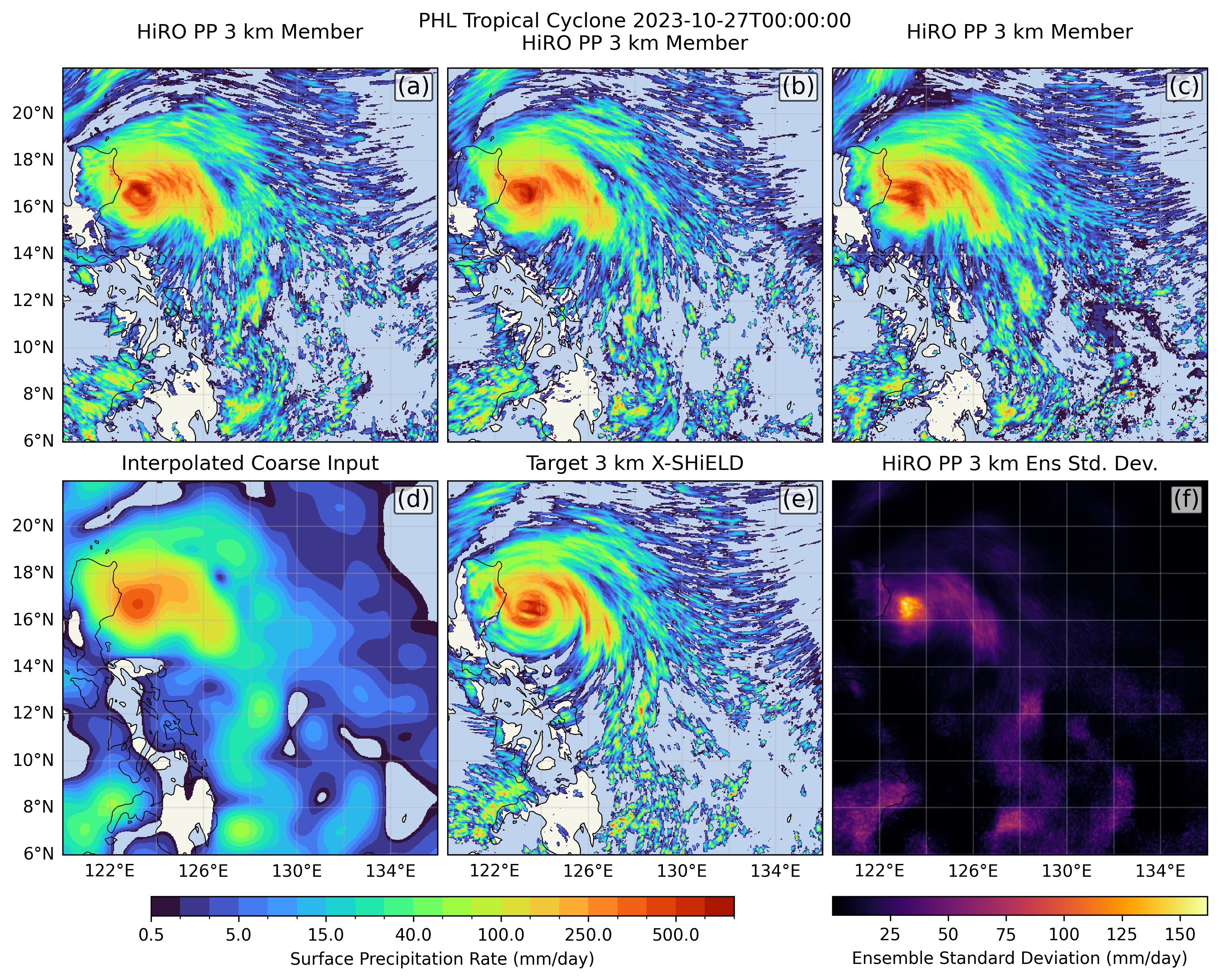}
     \caption{Perfect prediction downscaling for a tropical cyclone event over the Philippines showing (a-c) three selected generated samples from a 128-member ensemble, (d) bicubic interpolation of the coarse input, (e) target X-SHiELD 3 km data, and (f) standard deviation of the full generated ensemble.}
     \label{fig:phl-hurricane-sample-pp}
 \end{figure}

Figure \ref{fig:phl-hurricane-sample-pp} shows a selected set of generated samples (out of a 128-member ensemble) for an X-SHiELD-simulated tropical cyclone over the Philippines that highlight the probabilistic nature of HiRO. Here, moist convection produces many precipitation features that are unresolved at the coarse 100 km resolution.   The  small but well-organized target cyclone with a concentric eyewall is making landfall along the northern island (Fig. \ref{fig:phl-hurricane-sample-pp}e).  The interpolated coarse precipitation field (Fig. \ref{fig:phl-hurricane-sample-pp}d) captures the gross cyclone structure but none of the textural details.  The three generated samples (Fig. \ref{fig:phl-hurricane-sample-pp}a-c) all show realistic features like spiral bands and linear organization east of the cyclone, but with different details.  As in the highlighted HiRO-ACE case from Section \ref{sec:overview-key-results}, none of the three samples (or any of the other 125 generated samples) exhibit perfect eyewall structure, and the third selected sample has an unphysical double maximum in precipitation at the cyclone center (Fig \ref{fig:phl-hurricane-sample-pp}c).  Given the low amount of information from the coarse input, there are a wide range of potential storms that could be associated with it, and we would not expect the downscaling model to perfectly reproduce the target X-SHiELD event.

However, we expect the precipitation statistics of these generated storms to encompass the target, and indeed they do.  We can calculate standard deviations of the precipitation maps from the generated ensemble (Fig. \ref{fig:phl-hurricane-sample-pp}f) and combine them with maps of the ensemble-mean precipitation to get maps of 95\% ($\pm 2\sigma$) confidence intervals.  The X-SHiELD target falls outside these confidence intervals at only 5\% of locations, randomly scattered across the plotted domain (not shown).  Similarly, from a histogram of 6-hourly precipitation at fine grid points within the plotted region (SI Fig. S8), we see that the single X-SHiELD event lies within the 95\% precipitation rate confidence interval of the ensemble of PP-generated histograms for this event.

\subsection{Time-mean precipitation biases in complex geography}
\begin{figure}
    \centering
    \includegraphics[width=\textwidth]{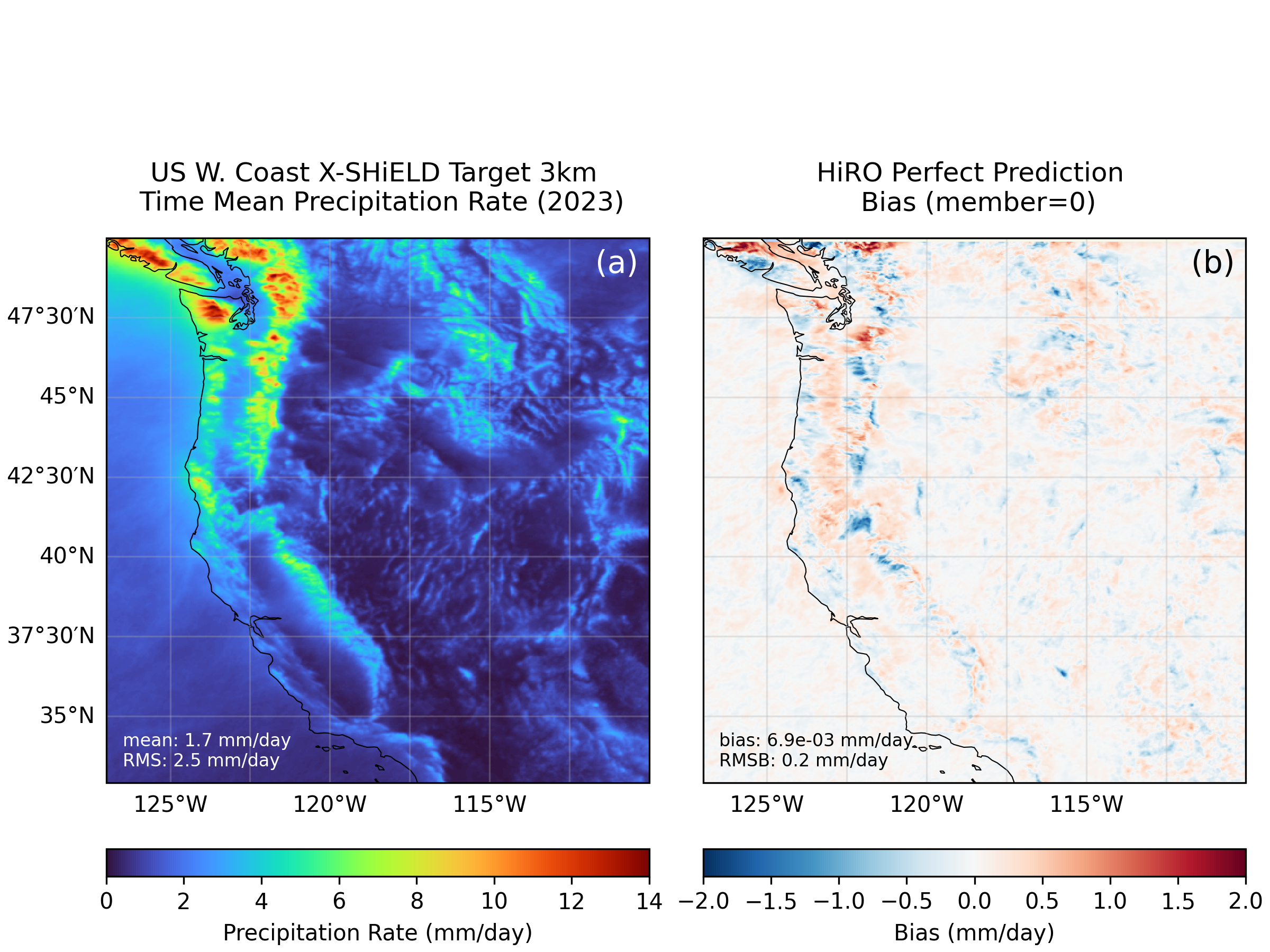}
    \caption{Time mean precipitation over the 2023 validation period of: (a) target X-SHiELD 3 km data over the west coast of the United States, (b) the bias of a single-member of the HiRO PP downscaled output ensemble. Regional mean metrics of the map area are given in the lower right for each panel. (See SI Fig. S9 for a map masked with 95\% significance threshold based on time-mean PP variability.)}
    \label{fig:wcoast-time-mean-pp}
\end{figure}

Finally, we show that PP downscaling has reassuringly small annual-mean biases compared to the 3 km X-SHiELD target over the 2023 hold-out year.  Figure \ref{fig:wcoast-time-mean-pp} displays biases over the western U.S., an example with complex geography.  We generated a 16-member HiRO PP downscaled ensemble over this region.  Figure \ref{fig:wcoast-time-mean-pp}a shows the target annual-mean precipitation over the held-out validation year of 2023, and Figure \ref{fig:wcoast-time-mean-pp}b shows the HiRO PP biases for a single ensemble member, which is an appropriate comparison with our single X-SHiELD target simulation.  The full ensemble is used to estimate 95\% location-specific significance thresholds for fine-scale internal variability of the ensemble mean in SI Figure S9.  

Figure \ref{fig:wcoast-time-mean-pp}b shows that the generated 3 km time mean precipitation is remarkably close to the target, with an RMSB of only 0.2~mm/day and virtually no area-mean bias.  This is much smaller than the RMSB of 0.8~mm/day for the bicubically interpolated coarse-grid precipitation (not shown), highlighting the added value of the ML residual correction. As expected, the PP bias tends to be larger over mountainous wet regions, but it is less than 1~mm/day almost everywhere.  The relative bias (i.e., bias as a fraction of target time-mean precipitation) is shown in SI Figure 9d.  The relative measure normalizes wet and dry regions, while significance thresholding ensures only relative biases rising substantially above this `noise' level should be interpreted as systematic errors of PP downscaling. This is most important where most precipitation falls in convective storm systems, e.g. the dry interior of the southwest U.S.  Significant relative biases suggest that HiRO tends to disperse precipitation away from topographic maxima into the drier lee-side zones. Excluding regions with time-mean precipitation less than 0.5~mm/day, the average absolute relative bias is around 7.5\%.

We did a single set of global PP generations covering the 2023 hold-out year, from which we computed annual-mean biases vs. the X-SHiELD simulation (SI Fig. S3a).  This confirms that HiRO PP (a single ML model trained on patches of global X-SHiELD data) is comparably skillful over diverse climates and geographic regions without any region-specific fine-tuning.  The largest biases are in tropical regions of convective activity where a single year time-mean of precipitation has large uncertainties due to km-scale internal variability. 

\section{HiRO-ACE: 100 km to 3 km downscaling of ACE2S}
\label{sec:hiro-ace-time-mean}

\begin{figure}
\centering
\includegraphics[width=\textwidth]{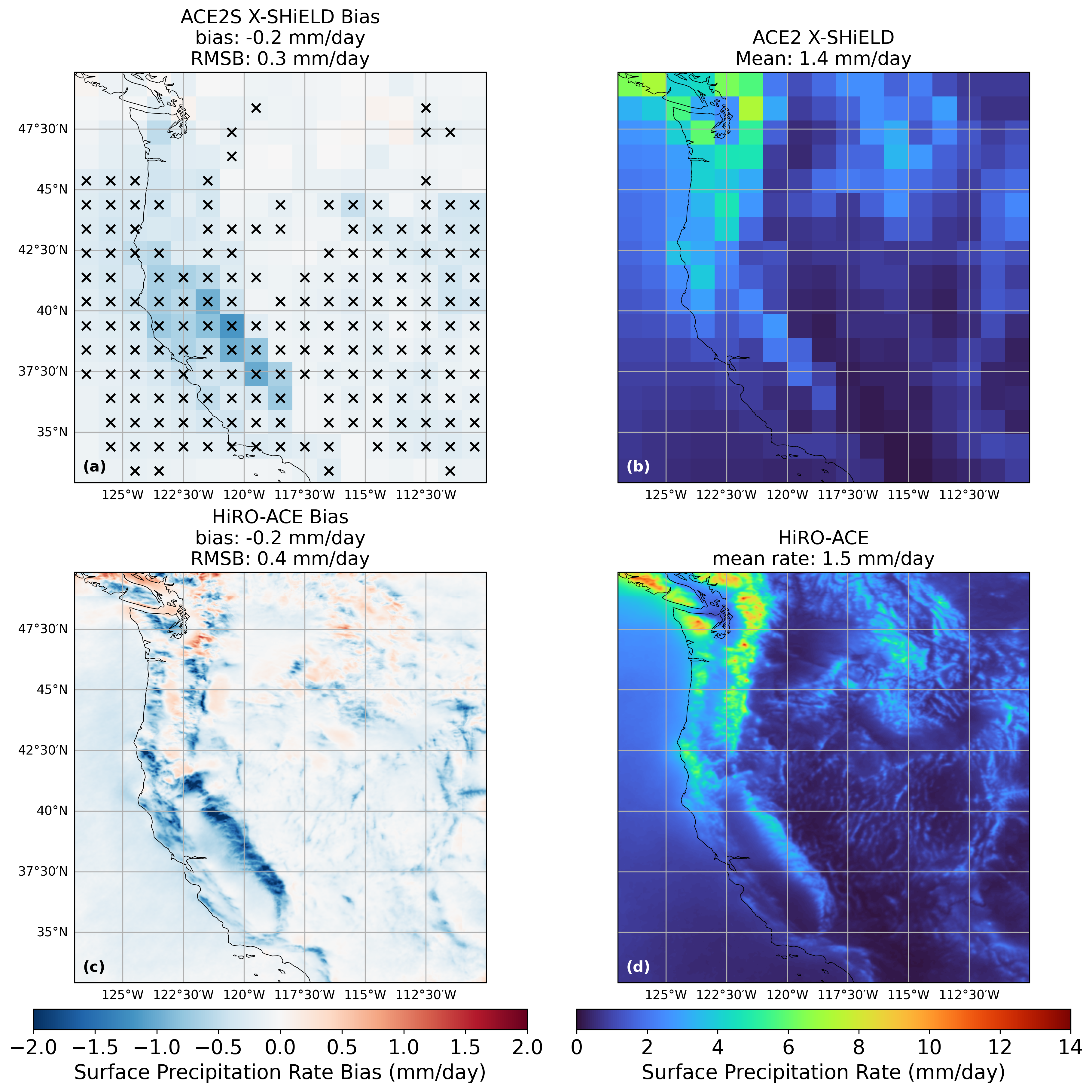}
\caption{(a) 10-year mean surface precipitation bias of ACE2S, run from 2014-2023, vs. X-SHiELD.  Crosses indicate gridpoints where ACE2S absolute biases are at least 0.1 mm/d and exceed 2 standard deviations of the noise floor due to unforced internal variability. (b) 10-year mean surface precipitation for ACE2S, (c) corresponding HiRO-ACE 10-year mean bias of surface precipitation, (d) 10-year mean HiRO–ACE surface precipitation.}
\label{fig:downscaled_western_us}
\end{figure}

Finally, to evaluate the end-to-end framework of HiRO-ACE, we downscale a 10 year run of ACE2S starting January 1, 2014. We downscale the 6-hourly output of the simulation over the Western U.S. region using a single HiRO realization. Figure~\ref{fig:downscaled_western_us} shows the time-mean surface precipitation fields for the coarse inputs (i.e. ACE2S) and the downscaled HiRO-ACE results. Generally, HiRO-ACE has low biases (absolute biases less than 0.5 mm/day, relative biases less than 10\%) for most of the Western U.S.  HiRO-ACE has a domain average RMSB of 0.4 mm/day, compared to the single-year HiRO PP RMSB of 0.2 mm/day. The areas with the largest biases, approaching 1 - 2 mm/day, are found over higher terrain in California, mainly the Sierra Nevada mountain range. Most of the HiRO-ACE bias is driven by time-mean precipitation underestimation in the southwestern U.S. by the ACE2S simulation (Fig. \ref{fig:downscaled_western_us}a), which was previously noted in Section \ref{sec:ace2s-results}.

% ACE2S slightly underestimates precipitation across a large portion of the domain. A `noise floor' analysis (see SI Text S3 for details) suggests much of the precipitation bias in this region is significantly (at 95\% confidence) above the range expected from internal variability (marked grid cells in Fig. \ref{fig:downscaled_western_us}a).  This could be due to small inconsistencies between the ERA5 data used to pre-train ACE2S and the 10-year X-SHiELD dataset used for fine tuning.

%Finally, to supplement Fig. \ref{fig:hurricane-focus-ace-hiro}, Fig. \ref{fig:event-histogram-comparison} presents regional precipitation histograms for three other weather events, comparing downscaled predictions of an ensemble of 6-hour HiRO-ACE forecasts vs. the X-SHiELD target. For all three events, two tropical cyclones and an atmospheric river over the Pacific Northwestern U.S., the performance of the HiRO PP and HiRO-ACE results are statistically indistinguishable even at the most extreme precipitation rates. 

To look at global behavior, we performed a 1-year single-member HiRO-ACE downscaling from \SI{65}{\degree S}--\SI{65}{\degree N} for the out-of-sample period (2023).  Time means and biases are shown in SI Figure S2 and S3. Overall, global-mean biases are negligible. HiRO-ACE displays larger pattern errors (due to underlying ACE2S biases) and RSMB (0.9 mm/day) compared to HiRO PP (RSMB of 0.3 mm/day), which is not surprising given the smaller averaging period.  Some coarse grid cell boundary imprinting is apparent in the equatorial east Pacific and Atlantic in both the HiRO PP and HiRO-ACE bias maps.

\section{Conclusions}
%Text here ===>>>
This work demonstrates HiRO-ACE, a machine-learned emulator for generating climate model outputs at 3km resolution. Our emulator consists of two model components: a 100 km coarse-resolution stochastic, autoregressive climate model emulator (ACE2S) and a corrective diffusion model (HiRO) which downscales the emulator’s output to 3 km resolution over almost any region worldwide. Both components of HiRO-ACE are trained on 9 years of the 3 km X-SHiELD global storm-resolving simulation (with a training dataset this short, ACE2S also benefits from pretraining on ERA5). 
%The coarse resolution emulator is trained using the previously-published ACE2 training methodology but two notable differences were required for producing low-bias precipitation outputs suitable for use as downscaling inputs: i) allow for a stochastic model by using a probabilistic loss function and ii) taking advantage of longer, complementary datasets by pretraining on ERA5 data. 

The ACE2S emulator accurately reproduces the statistics of the coarsened X-SHiELD dataset that it is tuned on. The one-year time mean of HiRO-ACE fine-resolution precipitation has relative biases of less than 10\% almost everywhere. In fact, these biases are statistically undetectable above internal weather noise in most locations (80\% of grid points from \SI{65}{\degree S}--\SI{65}{\degree N}). Notable significant coarse-grid precipitation biases due to ACE2S occur mainly over the tropical oceans, in subtropical/polar dry regions and sparse regions over land. Detectable fine-grid biases due to HiRO occur due to slight over-spreading of precipitation in some regions from wet mountain tops to adjacent dry lowlands.

Prediction snapshots of individual extreme precipitation events demonstrate HiRO-ACE can generate plausible km-scale samples of tropical cyclones, fronts, and convective events from a distribution of coarse-resolution inputs. Cumulative global precipitation histograms over a simulated year capture the extreme tail of the original X-SHiELD precipitation distribution.

This framework allows us to autoregressively generate coarse-resolution ensembles of multiyear time series and sub-select ranges of time and space to downscale to higher resolution. This output format is familiar to those in the broader community who are already using physical models for forecasting and applications to downstream impacts, but is generated hundreds of times faster than by dynamical downscaling. By construction, both ACE2S and HiRO are probabilistic, allowing users to generate large ensembles that account for weather and storm-scale variability across coarse and fine spatial scales. %Generating a single 3 km global rollout over one simulated year uses over 100 times less energy than an equivalent simulation using the numerical X-SHiELD model. Most users will probably only want to downscale small regions rather than the full globe, so the energy savings over an equivalent storm-resolving numerical model in those cases would increase by another factor of 10-100, depending on the region size. 

As in previous ACE2 studies, the framework can in the future be extended for operation in changing climates \cite{clark2025} or for coupling with other Earth system components \cite{duncan2025samudrace}. This extension will mainly be paced by the availability of future global km-scale simulations in multiple climates of sufficient length to train ACE2S (at least a decade and ideally several decades). 

Future versions of HiRO-ACE will be trained to downscale other variables of interest (e.g., surface temperature, snow-related fields, and surface winds) using a multi-variate extension of HiRO, similar to other downscaling models such as CorrDiff \cite{mardani_residual_2024}. HiRO-ACE or its individual components could be used, with only slight modification, in a weather forecasting setting. Lastly, ongoing advances in diffusion modeling \cite<e.g.,>{swift,zhou2024} could further speed up HiRO downscaling by potentially an order of magnitude.

\appendix

\section{ACE2S Loss Function}
\label{eq:crps-definition}

CRPS \cite{gneiting07} is defined as
\begin{align*}
\mathrm{CRPS}(F, x)
= - \int_{-\infty}^{\infty} \bigl(F(y) - \mathbf{1}\{y \ge x\}\bigr)^2 \, dy
\end{align*}
where $F$ is the cumulative distribution function for values in $\mathbb{R}$, and is shown to have an equivalent form of
\begin{align*}
    \text{CRPS}(F, y) \;=\; \mathbb{E}_{X \sim F}\!\left[\,|X - y|\,\right]
\;-\; \tfrac{1}{2}\,\mathbb{E}_{X, X’ \sim F}\!\left[\,|X - X’|\,\right],
\end{align*}
where $F$ is the random distribution function of values in $\mathbb{R}$ and $y \in \mathbb{R}$ is a true sample.

``Almost fair" CRPS \cite{lang24} is taken as
\begin{align*}
    \text{afCRPS}_{\alpha,M}(F, y) \;=\; \mathbb{E}_{X \sim F}\!\left[\,|X - y|\,\right]
\;-\; (1 - \frac{1 - \alpha}{M})\tfrac{1}{2}\,\mathbb{E}_{X, X’ \sim F}\!\left[\,|X - X’|\,\right],
\end{align*}
where $\alpha$ is a chosen parameter, and $M$ is the size of the ensemble. 

When we apply ``almost fair" CRPS at each grid point and then average across the spatial dimensions, we get 
\begin{align*}
    \overline{\text{afCRPS}_{\alpha,M}(F, y)} \;=\; \mathbb{E}_{X \sim F}\!\left[\,\frac{1}{N}\sum_{i=1}^N|X_i - y_i|\,\right]
\;-\; (1 - \frac{1 - \alpha}{M})\tfrac{1}{2}\,\mathbb{E}_{X, X’ \sim F}\!\left[\,\frac{1}{N}\sum_{i=1}^N|X_i - X_i’|\,\right],
\end{align*}
where subscript $i$ indicates the index of the grid point and $N$ is the total number of grid points.

Energy score \cite{gneiting07} is given by
\begin{align*}
    \text{ES}(F, \vec{y}) \;=\; \mathbb{E}_{\vec{X} \sim F} \big[ \, \lVert \vec{X} - \vec{y} \rVert \,\big]
\;-\; \tfrac{1}{2} \, \mathbb{E}_{\vec{X}, \vec{X}’ \sim F} \big[ \, \lVert \vec{X} - \vec{X}’ \rVert \,\big]
\end{align*}
and is defined for $y \in \mathbb{R}^n$. We apply the energy score independently to each complex coefficient of the spectral transform of the data, treated as vectors in $\mathbb{R}^2$, and then average it across these spectral coefficients. In this case the norm $\lVert \vec{x}\rVert$ is equivalent to the magnitude of the complex number it represents. This can be written as

\begin{align*}
    \overline{\text{ES}(\text{SHT} \circ F, \text{SHT}(y))} \;=\; \mathbb{E}_{\vec{X} \sim {\text{SHT} \circ F}} \big[ \, \frac{1}{M}\sum_{i=1}^M\lVert \vec{X}_i - \text{SHT}(y)_i \rVert \,\big]
\;-\; \tfrac{1}{2} \, \mathbb{E}_{\vec{X}, \vec{X}’ \sim \text{SHT} \circ F} \big[ \, \frac{1}{M}\sum_{i=1}^M\lVert \vec{X}_i - \vec{X}’_i \rVert \,\big]
\end{align*}
where subscript $i$ indicates the index of the spectral coefficient, $M$ is the total number of spectral coefficients, and the values at each spectral coefficient (for example $X_i$ and $SHT(y)_i$) are taken as vectors in $\mathbb{R}^2$, with double lines indicating the L1-norm.  This can be regarded as a `spectral CRPS'.

Our loss function is given by

\begin{align*}
    \mathcal{L}(F, y) = 0.9 \cdot \overline{\text{afCRPS}_{0.95, 2}(F, y)} + 0.1 \cdot \frac{2}{l}\overline{\text{ES}(\text{SHT} \circ F, \text{SHT}(y))}
\end{align*}

where SHT is the spherical harmonic transform, with its complex-valued outputs treated as vectors in $\mathbb{R}^2$ for the purpose of computing the energy score, $l$ is the spectral cutoff wavenumber for triangular truncation chosen as the number of latitude bands, and expected values are calculated with an ensemble size of 2. Our decision to take the average value of CRPS without area-weighting best matches the choice of MSE loss without area-weighting for ACE2.

We did not attempt using fair CRPS ($\alpha = 1$), although this should be permissible due to the inclusion of the spectral loss term, which constrains the target even in the case of a perfectly accurate prediction at one point.

\section{Additional HiRO Information}
\label{sec:hiro-supplement}

We train the HiRO denoiser using the standard score-matching framework described in \citeA{Karras2022ElucidatingTD}, with the log-normal distribution ($p_\sigma$) with values $P_{mean} = -1.2$ and $P_{std} = 1.8$.  We slightly increased  $P_{std}$ compared to previous work \cite<e.g.,>{brenowitz_climate_2025} to ensure that the training generates sufficient noise during training to fully saturate the target residual fields, even for high variance events such as tropical cyclones. 
 
 The diffusion denoising model is a 30M parameter Song U-Net \cite{Song2020ScoreBasedGM}, with hyperparameters adjusted for our 32:1 downscaling ratio and our patch size of $16 \times 16$ coarse grid cells. We use 128 channels, 7 levels, each with a single U-Net block, a base channel embedding multiplier of 6 and per-level multipliers of [1, 2, 2, 2, 2, 2, 2].  For inference, we use the standard Elucidated Diffusion Model (EDM) algorithm \cite{Karras2022ElucidatingTD} with $\sigma_{max} = 200$ (maximum noise) and $S_{churn}=0$. 
 
\subsection{Patch-based training and inference}
\label{app:hiro-patching}

As in cBottle \cite{brenowitz_climate_2025}, we train HiRO using a patch-based strategy to limit the GPU memory usage during generation and reduce the time to answer for downscaling specific regions of interest. For each batch, we pick a randomly shifted start point in the first $16^\circ \times 16^\circ$ patch of the global grid.  From that shifted starting point we decompose the domain of the loaded global data batch into patches, dropping any partial patches along the edges.  The random shift trains the model across a variety of starting configurations of the input data and fine topography, allowing for inference to use an arbitrary patch while maintaining performance.  Each collection of patches in the batched data is shuffled and run through the optimization procedure.  One drawback of this strategy is that it will undersample the edges of the domain, something to be considered if important and rare phenomena to be sampled for the corrective downscaling are along those edges.

We tested an ablation using a smaller input patch size of $8^\circ \times 8^\circ$ which resulted in slightly higher time-mean RMSBs over topography. We also tested training using a fixed domain patch decomposition, which yielded similar time-mean RMSBs when inference evaluation used the exact fixed patches from training.  However, performance degraded when evaluating downscaled patches shifted from those training positions.

For inference, if the target domain to downscale is a single $16^\circ \times 16^\circ$ patch, we just use the diffusion model to generate realizations of the desired field.   If the target domain to downscale encompasses more than one patch, we cover it with a set of $16^\circ \times 16^\circ$  patches with a single coarse grid cell ($1^\circ$~width) of overlap. We generate each individual patch, average the generations in the overlap region, and crop the final output to the requested domain.  This strategy prevents large discontinuities at patch edges when generating larger snapshots, at the expense of some over-smoothing and distortion in the overlap region (32 grid cells at 3 km), which can show up in 2nd-order statistics calculated on the output fields.  This is a simple, effective, but less sophisticated version of the ``multidiff'' strategy used by cBottle.

\subsection{Best checkpoint selection during training}

The best checkpoint is selected to minimize the lowest absolute normalized bias in the tail for fully denoised precipitation histogram beyond the 99.99th percentile of the validation dataset target, calculated as
\begin{equation}
     \frac{1}{N} \sum_{\rho_i>\rho_\mathrm{target99.99}} \left | \frac{\rho_\mathrm{pred, i}}{\rho_\mathrm{target, i}} -1 \right |.
\end{equation}
$N$ is the number of bins with precipitation rate greater than the 99.99th percentile of target density $\rho_{\mathrm{target99.99}}$. $\rho_i$ is the bin density of histogram bin $i$. Using this metric for checkpoint selection improves resulting model extremes while giving similar time-mean skill as the continuous ranked probability score (CRPS). The validation set consists of 76 global snapshots from the held-out year of data. This metric is calculated every 10 denoiser training epochs as it is time-consuming.

%%%%%%%%%%%%%%%%%%%%%%%%%%%%%%%%%%%%%%%%%%%%%%%
% Optional Glossary, Notation or Acronym section goes here:
%
% Glossary is only allowed in Reviews of Geophysics
%  \begin{glossary}
%  \term{Term}
%   Term Definition here
%  \term{Term}
%   Term Definition here
%  \term{Term}
%   Term Definition here
%  \end{glossary}

%%%%%%%%%%%%%%%%%%%%%%%%%%%%%%%%%%%%%%%%%%%%%%%
% Acronyms
%% NOTE that acronyms in the final published version will be spelled out when used in figure captions.
%   \begin{acronyms}
%   \acro{Acronym}
%   Definition here
%   \acro{EMOS}
%   Ensemble model output statistics
%   \acro{ECMWF}
%   Centre for Medium-Range Weather Forecasts
%   \end{acronyms}

%%%%%%%%%%%%%%%%%%%%%%%%%%%%%%%%%%%%%%%%%%%%%%%
% Notation
%   \begin{notation}
%   \notation{$a+b$} Notation Definition here
%   \notation{$e=mc^2$}
%   Equation in German-born physicist Albert Einstein's theory of special
%  relativity that showed that the increased relativistic mass ($m$) of a
%  body comes from the energy of motion of the body—that is, its kinetic
%  energy ($E$)—divided by the speed of light squared ($c^2$).
%   \end{notation}

%%%%%%%%%%%%%%%%%%%%%%%%%%%%%%%%%%%%%%%%%%%%%%%
%
% DATA SECTION and ACKNOWLEDGMENTS
%
%%%%%%%%%%%%%%%%%%%%%%%%%%%%%%%%%%%%%%%%%%%%%%%

\section*{Open Research Section}
The code for training ACE2S and HiRO and inference is available at \newline https://github.com/ai2cm/ace/tree/v2026.1.1.  Model weights, forcing data, initial conditions, and configurations to generate example HiRO-ACE outputs over the 2014--2023 period are available at https://huggingface.co/allenai/HiRO-ACE. The processed ERA5 dataset used to pre-train ACE2S is available on a public requester-pays Google Cloud Storage (GCS) bucket at gs://ai2cm-public-requester-pays/2024-11-13-ai2-climate-emulator-v2-amip/data/era5-1deg-1940-2022.zarr (about 1.5TiB). The coarsened 10-year X-SHiELD dataset and 3 km precipitation dataset (about 5 TiB) will be published prior to publication.  The analysis and plotting code with intermediate data for plot reproduction are available at https://zenodo.org/records/18475758.

\section*{Conflict of Interest disclosure}
“The authors declare there are no conflicts of interest for this manuscript.”

\acknowledgments
Ai2 is supported by the estate of Paul G. Allen. We thank the NOAA Geophysical Fluid Dynamics Laboratory for performing the X-SHiELD simulation and supporting our use of its output, which enabled this work; in particular we appreciate Alex Kaltenbaugh for carrying out the initial post-processing of the data, which facilitated its use in downstream applications, including our workflow, and Linjiong Zhou and Kai Cheng for their contributions to X-SHiELD development. We acknowledge ECMWF for generating and providing the ERA5 dataset used for pretraining. We thank Noah Brenowitz and Mike Pritchard for helpful discussions about the application of diffusion modeling for downscaling. This research used resources of NERSC, a U.S. Department of Energy Office of Science User Facility located at Lawrence Berkeley National Laboratory, using NERSC award BER-ERCAP0030562.  We acknowledge the use of AI tools to assist with this research (code generation) and prepare the manuscript (writing review, stylistic suggestions, and drafting of summary abstracts).  The models used include Google Gemini 2.5 Flash and 2.5 Pro; OpenAI GPT-4o and o1; and Anthropic Claude Sonnet 4, Claude Sonnet 4.5, and Claude Opus 4.5.

%%%%%%%%%%%%%%%%%%%%%%%%%%%%%%%%%%%%%%%%%%%%%%%
% REFERENCES and BIBLIOGRAPHY
%
% \bibliography{<name of your .bib file>} don't specify the file extension
% don't specify bibliographystyle
%
%%%%%%%%%%%%%%%%%%%%%%%%%%%%%%%%%%%%%%%%%%%%%%%

%Reference citation instructions and examples:
%
% Please use ONLY \cite and \citeA for reference citations.
% \cite for parenthetical references
% ...as shown in recent studies (Simpson et al., 2019)
% \citeA for in-text citations
% ...Simpson et al. (2019) have shown...
%
%
%...as shown by \citeA{jskilby}.
%...as shown by \citeA{lewin76}, \citeA{carson86}, \citeA{bartoldy02}, and \citeA{rinaldi03}.
%...has been shown \cite{jskilbye}.
%...has been shown \cite{lewin76,carson86,bartoldy02,rinaldi03}.
%... \cite <i.e.>[]{lewin76,carson86,bartoldy02,rinaldi03}.
%...has been shown by \cite <e.g.,>[and others]{lewin76}.
%
% apacite uses < > for prenotes and [ ] for postnotes
% DO NOT use other cite commands (e.g., \citet, \citep, \citeyear, \nocite, \citealp, etc.).
%

\section{Supplementary Information}

\noindent\textbf{Table of Contents}
%%%Remove or add items as needed%%%
\begin{enumerate}
\item Text S1 to S4
\item Figures S1 to S9
\end{enumerate}

In the supplementary materials we cover sensitivities of design and hyperparameter choices in Section S1.  The determination of the weighting of the nodal and spectral terms in the CRPS loss is covered in Section S2.  In Section S3, we give a description of the time mean noise floor calculations for ACE2S and HiRO.  And finally, we describe ACE2S tropical cyclone characteristics in Section S4.

%\clearpage

%Delete all unused file types below. Copy/paste for multiples of each file type as needed.
\noindent\textbf{S1. Sensitivities of ACE2S to key design and hyperparameter choices}

We did not do any hyperparameter optimization of the ACE2S model architecture (e.g., model embedding size, per-variable weighting, number of SFNO blocks), using values chosen for ACE2 by \citeA{wattmeyer25}. 
The noise channel count of 64 was chosen because it worked well using a different dataset and training scheme, but was not optimized for this dataset and training scheme.

\noindent\textbf{S1.1 Weighting of nodal and spectral CRPS loss terms}

Using nodal CRPS alone produced poor results, with excessive small-scale variability similar to \citeA{lang24}. We tried weighting at 90\%/10\% and 10\%/90\% ratios for the nodal and spectral losses. Time-mean climate biases and short-term weather skill were insensitive to this choice. The spectral-heavy 10\%/90\% weighting led to speckles of small positive precipitation during inference in regions without precipitation in the reference data, causing a large relative time-mean bias in desert regions.  It also led to spectral ringing artifacts in some variables. Our choice of 90\%/10\% loss weighting removed both these artifacts, at the expense of some excess small-scale spectral power in stratospheric variables.  %The CRPS-heavy weighting was used due to limitations in available compute resources, an equal weighting shows some improvements to spectral power without the issues of the spectral-heavy loss.

\noindent\textbf{S1.2 Comparison of ACE2S multi-step rollout training method with alternatives}
%We pre-trained ACE2S on ERA5 for NEED epochs, following \cite{wattmeyer25}.

We compared the chosen multi-step rollout for the second stage of ACE2S training (using last-step training on a range of look-ahead periods with empirically chosen probabilities) with some other plausible alternatives. These included:  the 2-step scheme used in ACE2, a scheme which accumulated the loss over 20 forward steps, and a scheme which used equal probability for 1 through 20 steps in a one-step loss. The 2-step scheme had increased inference error even after more epochs of training. Using equal 1-20 step probability led to large errors in single-step evolution of stratospheric variables. In early tests which used ERA5 for both stages, the runs averaging over 20 timesteps perform as well as the chosen scheme, but with significantly increased computational cost, taking 7 times longer per batch than the final configuration for the X-SHiELD finetuning. This is due both to computing additional forward steps, and because in this case training must backpropagate over each forward step and must always compute 20 forward steps.

\noindent\textbf{S1.2 Noise conditioning}

We tested conditioning on Gaussian noise uncorrelated between Gaussian grid points as an alternative to spherically isotropic noise. This led to uncorrelated noise-like small-scale variability in polar regions during the initial 1-step ERA5 training stage. While this variability is not represented in spectral power, did not accumulate, and did not cause large errors, spherically isotropic Gaussian noise removed this artifacting without downsides. %The energy score loss was not able to correct these biases as they are present in grid scales not represented by the spectral transform of the data.

\noindent\textbf{S2 Sensitivity of ACE2S to ERA5 Pretraining }
\label{sec:ace_ablations}

We tried pretraining with a 1-step loss using the 10 years of X-SHiELD outputs in place of our 44-year ERA5 dataset. While simpler, this substantially worsened the climate biases on ACE2S (Fig. ~\ref{fig:climate_bias_panel}), including a nearly 50\% increase in the RMSB of surface precipitation.  Presumably, the X-SHiELD dataset is shorter than ideal for training an accurate stochastic climate emulator, and ERA5 pretraining helps address this. We obtained nearly as good results when pretraining using climate model output, e.g. a SHiELD AMIP simulation from 1940-2021 with $1^\circ \times 1^\circ$ grid resolution, in place of ERA5.  Such pretraining outputs could be produced for multiple climates and might be more suitable for training an emulator of a km-scale model that is accurate across a range of climates.

For spherical power spectrum and the distribution of surface precipitation rates, we find that doing the 1 - step training on ERA5 instead of X-SHiELD does not make a significant improvement.

\noindent\textbf{S3 Noise Floor Analysis}
\label{sec:noise-floor}

A critical part of our model evaluation is estimating the  spatial biases of HiRO, ACE2S, and HiRO-ACE in predicting time-mean precipitation of the reference km-scale model, using a limited number of simulations of finite length.  In particular, do those biases rise significantly above levels expected from random atmospheric variability, and if so, where?  In this section, we present some simple statistical analysis to this end based on `noise floor' estimation, also used by \citeA{Dun2024}, \citeA{BlochJohnson2024TheGF}, \citeA{Wu2025}, and others.

\noindent\textbf{S3.1 Noise floor for a general 2D field}

We start by deriving formulas using annual means of a general field $F$ predicted by a multimember ensemble generated by ACE2S and/or downscaled by HiRO, for which we have a single comparable km-scale reference simulation. We then handle some additional issues specific to our application to precipitation downscaling.

Suppose we have an ensemble of $M$ identically-forced simulations of the global atmosphere (i.e. they have the same time history of SST, sea-ice, greenhouse gases, etc.) Consider an atmospheric field $F(x,y,t)$ that is predicted by these simulations for $N$ years, defined at a globally distributed set of grid points $(x,y)$.  We denote the time series of its annual-mean values in ensemble member $m$ and year $n$ as $\hat{F}_m(x,y,n)$.  We regard this as a sample of a random variable with a true ensemble mean $\overline{F}(x,y,n)$  (which includes the response of $F$ to time-varying forcing), and we define the anomaly of ensemble member $m$ in year $n$ (which isolates internal unforced atmospheric variability) as
\begin{equation}
  \delta \hat{F}_m(x,y,n) =  
    \hat{F}_m(x,y,n) - \overline{F}(x,y,n).  
\end{equation}
We assume that anomalies are independent and identically distributed across different years and ensemble members.  This is reasonable because (1) the dominant timescales of internal weather variability are much faster than a year, decorrelating ensemble member anomalies from each other and between years, and (2) the statistics of internal atmospheric variability are only modestly dependent on year-to-year forced changes in the mean atmospheric state.

With this key assumption, we can define a 1-year location-dependent `noise floor' $\sigma_F(x,y)$ as the true standard deviation of anomalies $\hat{F}_m(x,y,n)$. To accurately estimate $\sigma_F(x,y)$ from a ensemble of simulations, we need multiple ensemble members; this estimate can be improved by using multiple simulated years.  For now, we will assume that we know $\sigma_F(x,y)$.

We wish to compare this ensemble with a single identically-forced reference simulation (with the expensive X-SHiELD global km-scale model, using historical 2014-2023 SSTs, in our case).  For evaluating ACE2S we use annual means from the reference output  coarsened to the ACE2S grid, $\hat{F}_{c,ref}(x,y,n)$.  For evaluating HiRO we use the corresponding fine-grid correction to this,
$\hat{F}_{f,ref}(x,y,n)$.  In either case, we analyze the `deviations'
\begin{equation}
   \Delta \hat{F}_m(x,y,n) 
   = \hat{F}_m(x,y,n)- \hat{F}_{s,ref}(x,y,n),
\label{eq:anom}
\end{equation}
where the spatial resolution '$s$' is either '$c$' (coarse) or '$f$' (fine).
From a single reference simulation, we cannot easily isolate the unforced variability of $\hat{F}_{s,ref}(x,y,n)$, but it is still formally useful to define a `true' reference mean $\overline{F}_{s,ref}(x,y,n)$, which could be constructed from a large ensemble of identically-forced reference simulations, if we could afford to do that.  We define the reference anomaly for the single reference simulation that we actually have:
\begin{equation}
  \delta \hat{F}_{s,ref}(x,y,n) =  
    \hat{F}_{s,ref}(x,y,n) 
    - \overline{F}_{s,ref}(x,y,n).   
    \label{eq:refanom}
\end{equation}

With this notation, we can decompose the deviations into a 'true' systematic error and random errors due to internal variability in the ensemble simulation and the reference simulation, respectively:
\begin{equation}
    \Delta \hat{F}_m(x,y,n) 
   = [\overline{F}_m(x,y,n) 
     - \overline{F}_{s,ref}(x,y,n)] 
     + \delta \hat{F}_m(x,y,n)  - \delta \hat{F}_{s,ref}(x,y,n).    
\label{eq:dev-decomp}
\end{equation}
We can estimate the PDFs of the two random error terms assuming they have Gaussian statistics.
By definition, $\delta \hat{F}_m(x,y,n)$ has a standard deviation equal to the noise floor $\sigma_F(x,y)$, so
\begin{equation}
  \delta \hat{F}_m(x,y,n) \sim n(0,\sigma_F).    
\end{equation}
Assuming the internal variability of the simulated ensemble is also a reasonable estimate of that of the reference simulation, then
\begin{equation}
  \delta \hat{F}_{s,ref}(x,y,n) \sim n(0,\sigma_F).  
\end{equation}
Furthermore, it is reasonable to assume that internal variability in the reference simulation is independent of that in the ensemble members.  

We now have the building blocks needed to construct 
uncertainty ranges for statistical significance of deviations of means of ensemble members from the reference data.  All three cases that we'll need to consider involve comparing the mean of some combination of $M \ge 1$ ensemble members with a single reference realization using a time-average of $N \ge 1$ years.  The resulting random uncertainty $\epsilon^{M,N}$ will have the PDF
\begin{equation}
  \epsilon^{M,N} \sim n\left(0,\sigma_F\left[\frac{1}{MN} + \frac{1}{N}\right]^{1/2}\right).  
\label{eq:epsMN}
\end{equation}
Using a 2 standard deviation (approximately 95\% 2-sided) confidence interval, we can infer from $M$ ensemble members and $N$ years of ensemble and reference data that the ensemble and time average $\hat{F}^{M,N}(x,y)$ is significantly different from that of $\hat{F}^N_{s,ref}(x,y)$ if 
\begin{equation}
   |\Delta \hat{F}^{M,N}(x,y)| 
   = |\hat{F}^{M,N}(x,y) - \hat{F}^N_{s,ref}(x,y)|
   > 2\sigma^{M,N}_F,
\label{eq:notnoiseMN}
\end{equation}
where the post-averaging noise floor is
\begin{equation}
   \sigma^{M,N}_F =\sigma_F
       \left[\frac{1}{MN} + \frac{1}{N}\right]^{1/2}.   
\label{eq:sigmaMN}
\end{equation}
Averaging over more years reduces the post-averaging noise floor more effectively than averaging over more ensemble members.

\noindent\textbf{S3.2 Application to ACE2S}

For ACE2S, we have an $M=10$ member ensemble of simulations for $N=10$ years. This is large enough to enable reliable estimates of the ACE2S 1-year noise floor $\sigma^{ACE2S}_P$ and hence an estimate of the 10-year single-ensemble member noise floor $\sigma_P^{1,10}$ used for the significance mask in Fig. 6. 

\noindent\textbf{S3.3 Application to HiRO}

For HiRO, we have an $M=16$ member PP ensemble generated over the western United States patch for the single year of 2023 ($N=1$).

\noindent\textbf{S4 Tropical Cyclone Climatology in ACE2S}
\label{sec:tc_ace}

Tropical cyclones (TCs) are an important target for downscaling because of their high impact and spatial compactness. In the HiRO - ACE framework, all of the information about a TC comes from the coarse resolution data, which during inference is generated by ACE2S. Using the TC tracking methodology outlined in \citeA{wattmeyer25}, we investigate the ability of ACE2S to reproduce the frequency, spatial distribution, and TC strength distribution compared to coarsened X-SHiELD (Figures \ref{fig:tc_climatology} and \ref{fig:tc_strength}). In general, ACE2S slightly overestimates the number of TCs in the major basins; the annual number of TCs is ~10 more per year than coarsened X-SHiELD. The deterministic ACE2 matches X-SHiELD more closely in TC locations and frequency. 

We also compared the histograms of maximum coarse-grid winds and minimum coarse-grid sea level pressure in TCs simulated by deterministic ACE2 and stochastic ACE2S. Overall, both ACE2S and ACE capture the PDFs of these quantities compared to coarsened X-SHiELD, although ACE2S tends to overly favor TC maximum wind speeds in the 25 - 30 m/s range, and neither ACE2 or ACE2S captures the most extreme coarsened TC winds simulated by X-SHiELD.

\clearpage

\begin{figure}[h!]
    \centering
    \includegraphics[width=\linewidth]{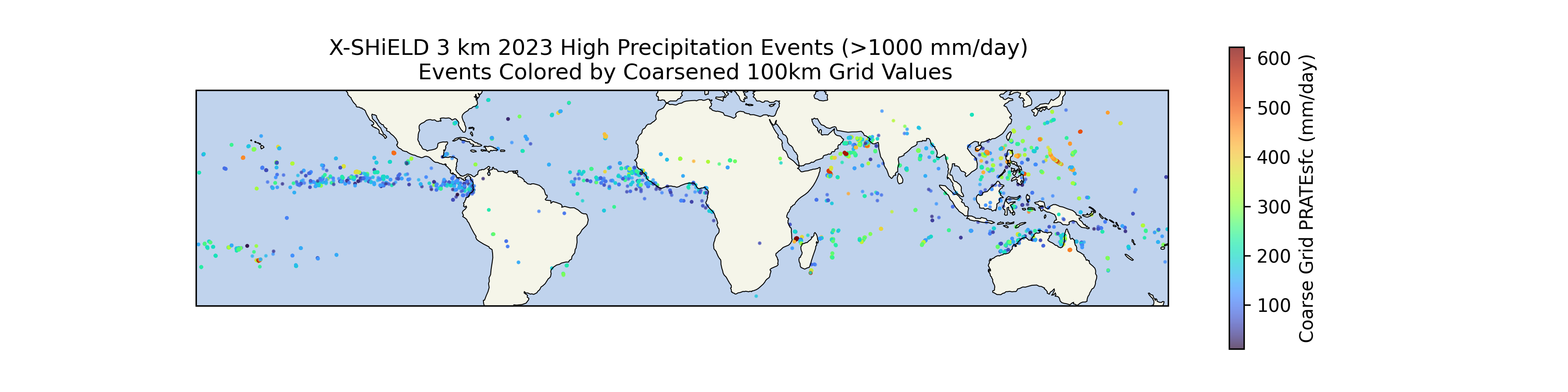}
    \caption{Locations of high precipitation rate events (
    $>$1000 mm/day) in the X-SHiELD 3 km target data for the 2023 time period.  Each point represents a single fine resolution grid cell exceedance event and is colored by the parent coarsened X-SHiELD grid cell (100 km) precipitation rate at the same time.}
    \label{fig:target-extreme-precip-events}
\end{figure}

\begin{figure}[h!]
\centering
\includegraphics[width=0.75\textwidth]{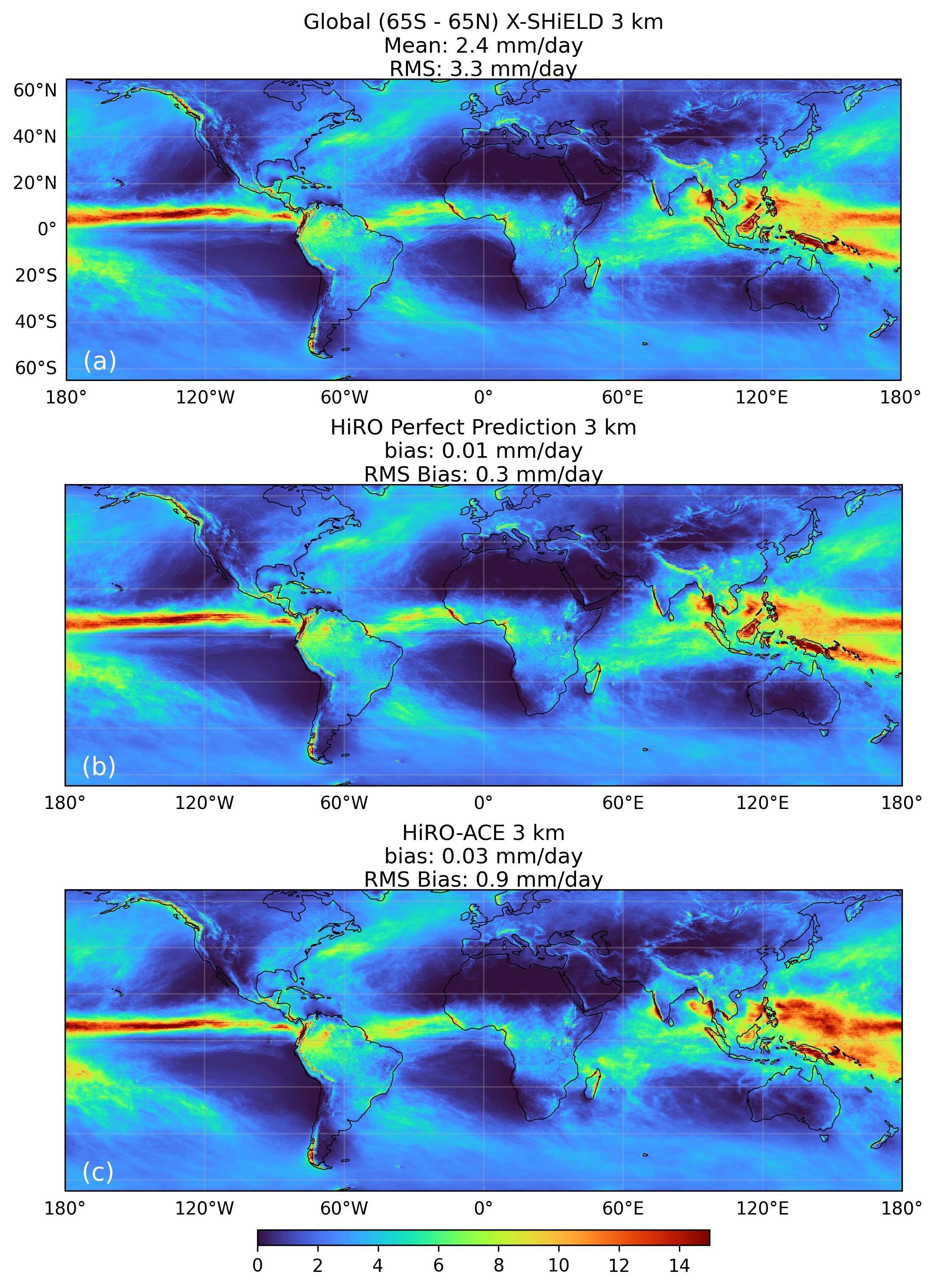}
\caption{Global time mean 6 h average surface precipitation rate (\SI{65}{\degree S}--\SI{65}{\degree N}) over the holdout validation period (2023) for (a) target X-SHiELD at 3 km, (b) a single ensemble member of HiRO perfect prediction downscaled data, and (c) ACE2S free running simulation over the same period downscaled by HiRO (single ensemble member).}
\label{fig:downscaled-global-means}
\end{figure}

\begin{figure}[h!]
\centering
\includegraphics[width=\textwidth]{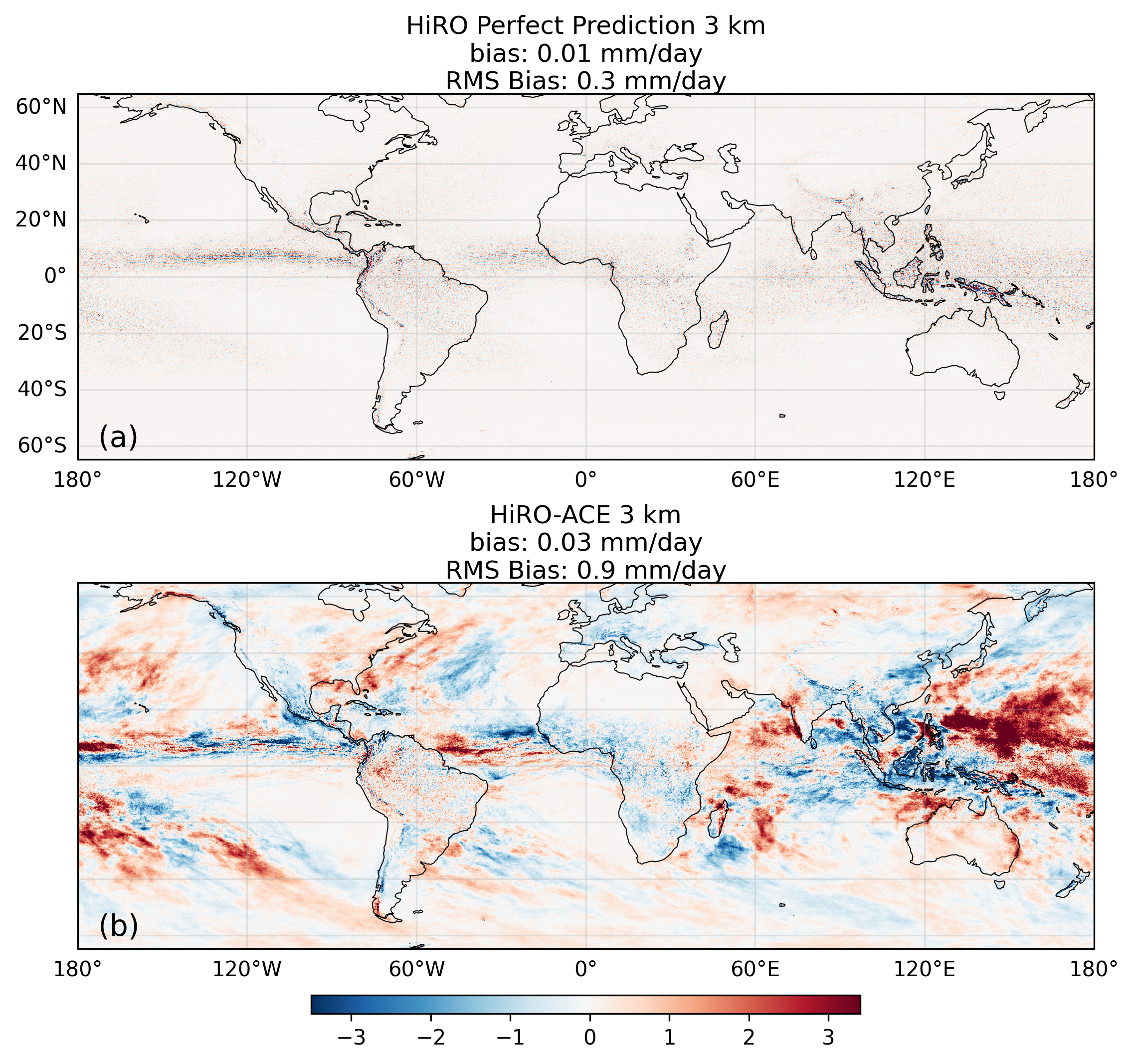}
\caption{As in Fig. \ref{fig:downscaled-global-means} but instead with biases plotted for (a) the HiRO perfect prediction data (b) the HiRO-ACE data.}
\label{fig:downscaled-global-means-bias}
\end{figure}

\begin{figure}[h!]
    \centering
    \includegraphics[width=0.75\textwidth]{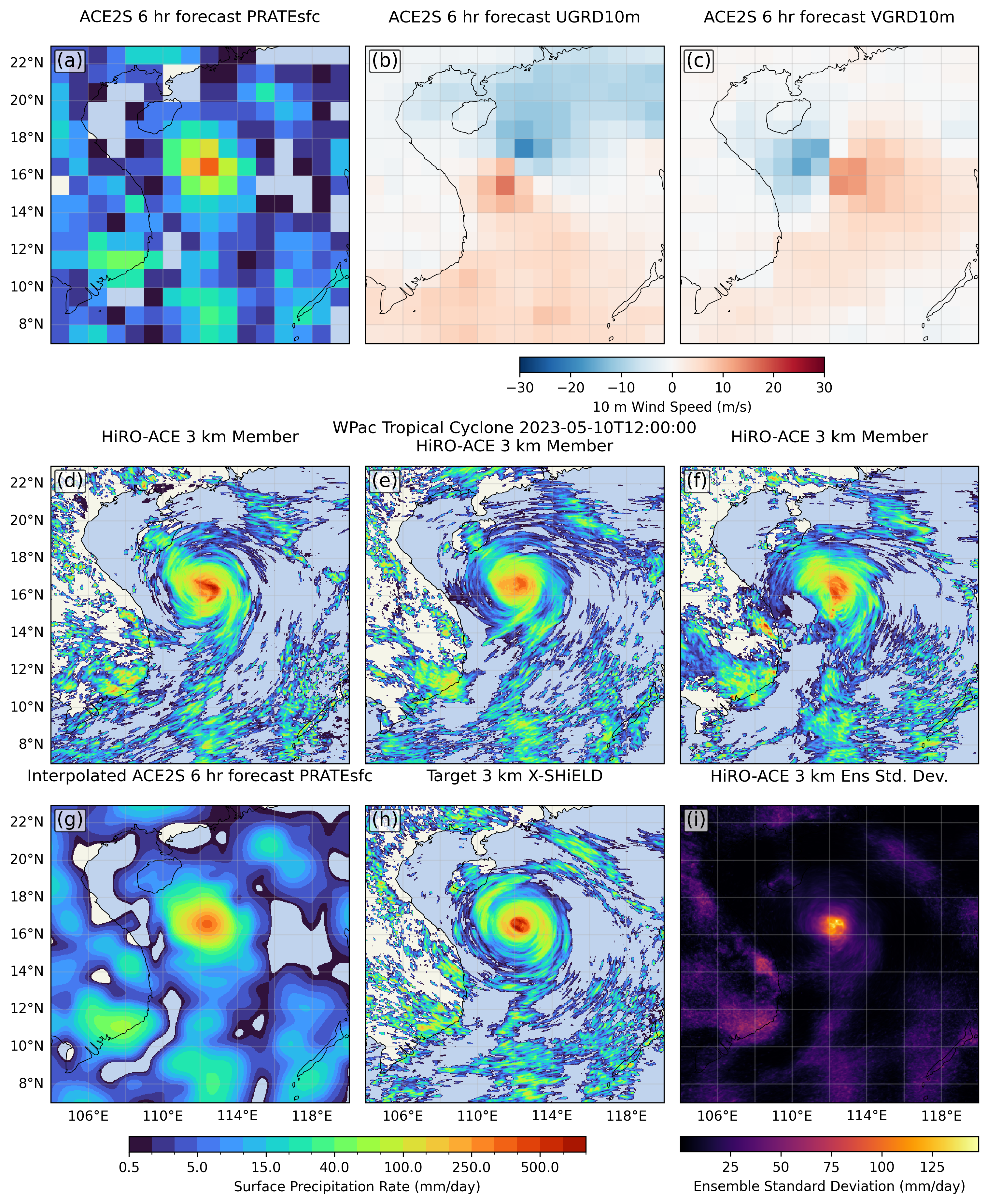}
    \caption{A HiRO-ACE downscaled event including an example of ACE2S (a--c) coarse 100 km fields, (d--f) selected generated 3 km samples, (g) an example of interpolated surface precipitation rate conditioning input, (h) X-SHiELD 3 km target, and (i) the generated ensemble standard deviation.  The ensemble generation uses 16 single step ACE2S forecasts (6 hours) initialized from the previous timestep and then uses HiRO to generate an 8-member ensemble for each forecast member to reach 128 total members.  }
    \label{fig:combined-wpac-hurricane-extended}
\end{figure}

\begin{figure}
\centering
\includegraphics[width=0.75\textwidth]{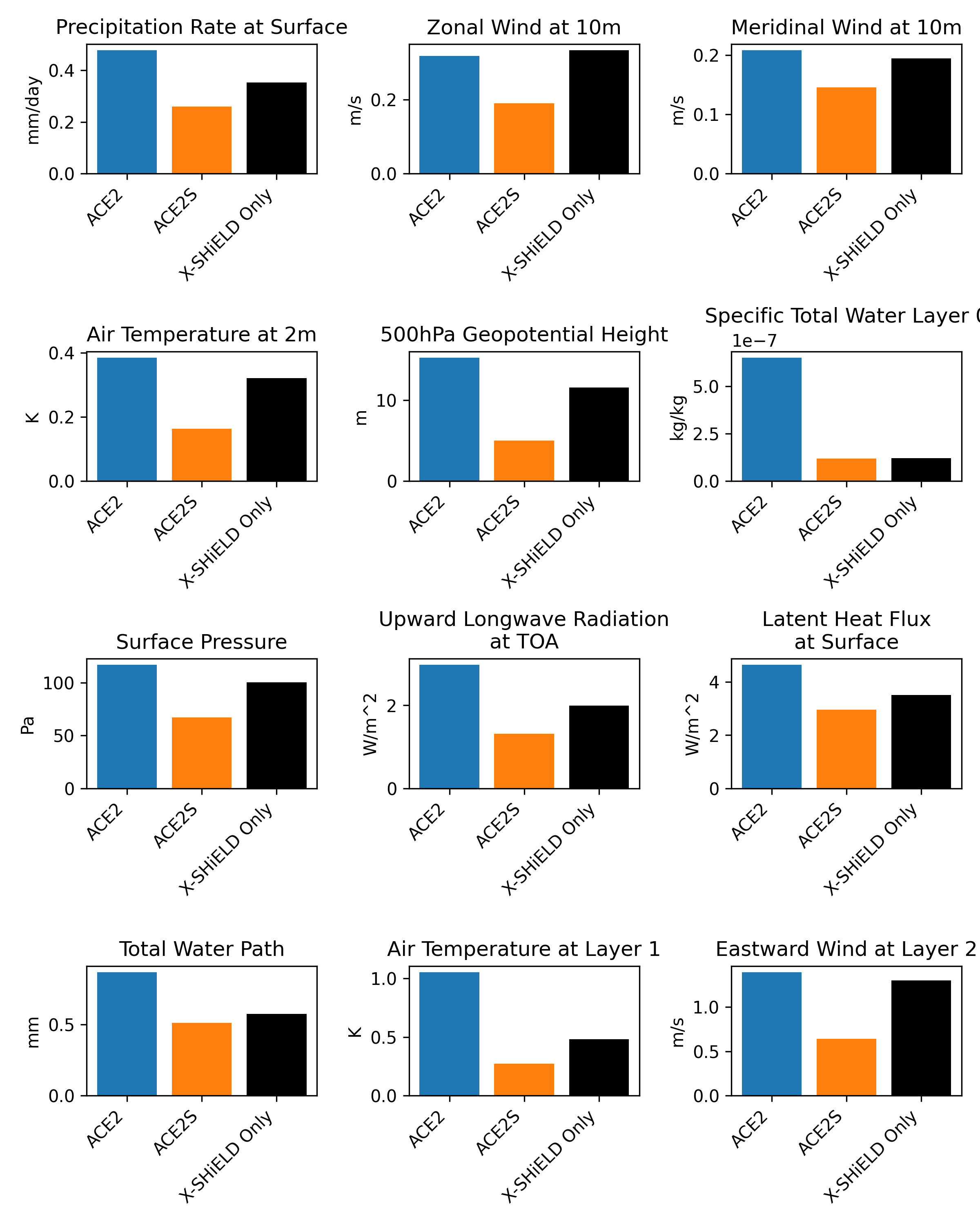}
\caption{10-year time-mean RMSE for ACE2 (blue), ACE2S pretrained on ERA5 (orange), and ACE2S without ERA5 pretraining inputs (black).}
\label{fig:climate_bias_panel}
\end{figure}

\begin{figure}
\centering
\includegraphics[width=\textwidth]{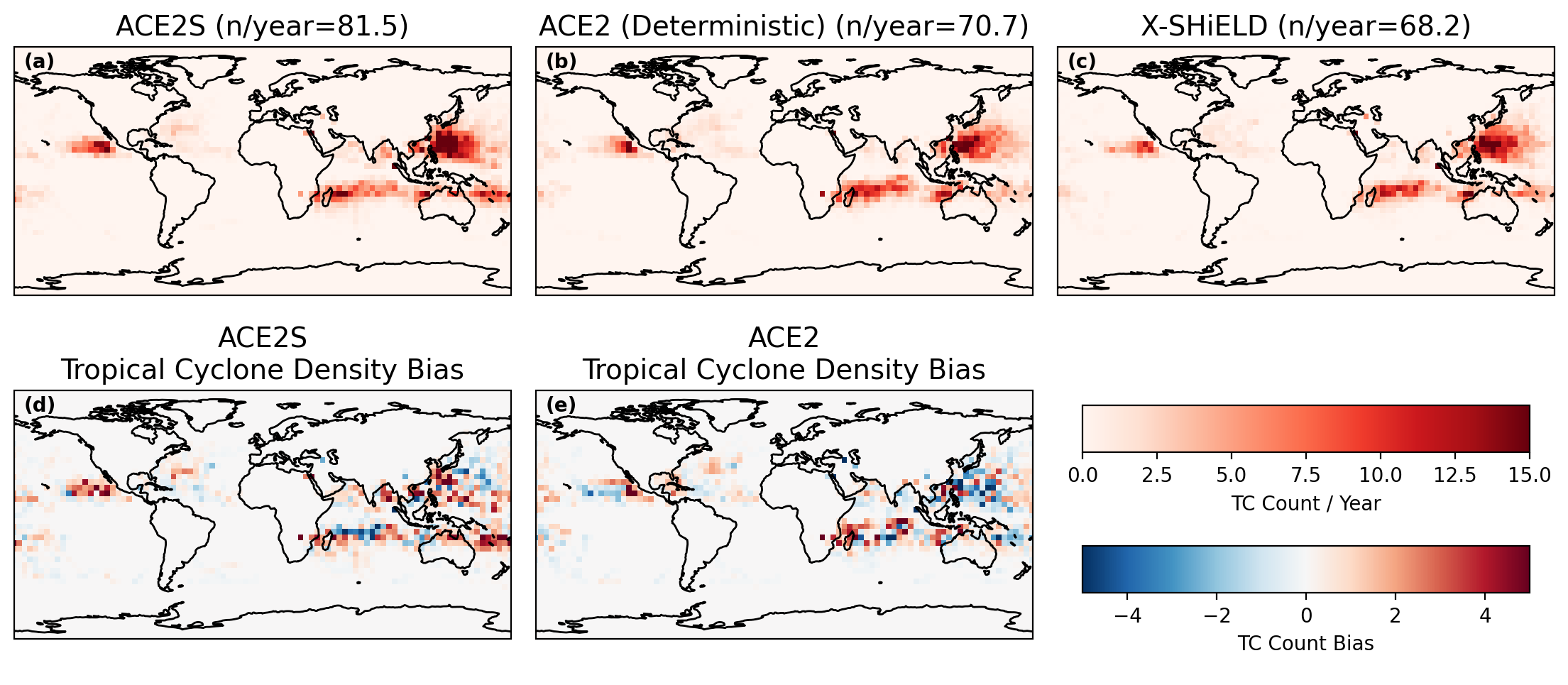}
\caption{10-year tropical cyclone density on a \SI{4}{\degree} by \SI{4}{\degree} grid for ACE2S (a), ACE2 X-SHiELD (b), and coarsened X-SHiELD (c). Number of cyclones per year are shown in for each model. Tropical cyclone density bias for ACE2S (d) and ACE (e). }
\label{fig:tc_climatology}
\end{figure}

\begin{figure}
\centering
\includegraphics[width=\textwidth]{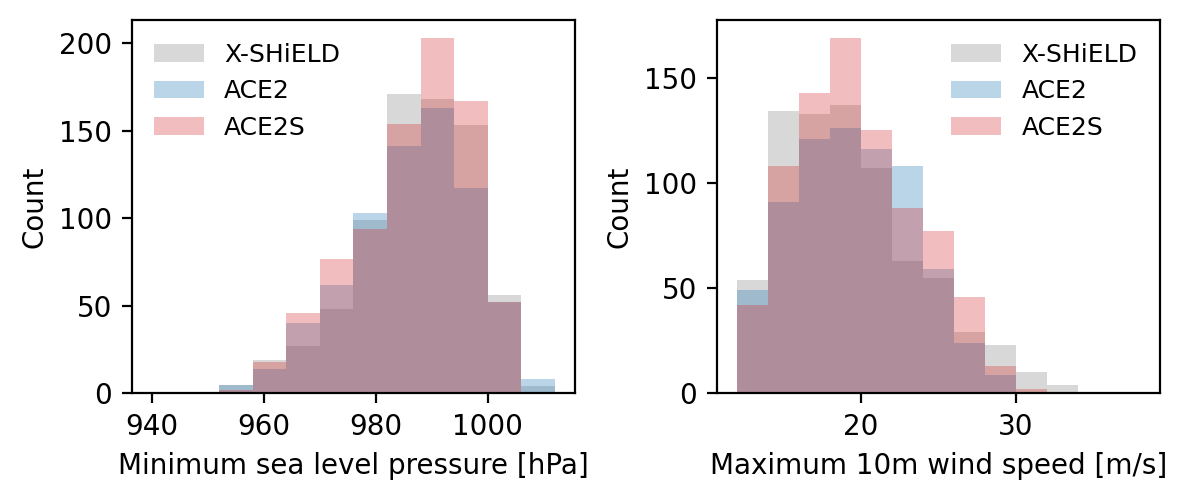}
\caption{The (left) minimum sea-level pressure and (right) maximum 10 meter wind speed within \SI{2}{\degree} of the sea-level pressure minimum across the 10-year period of Fig. \ref{fig:tc_climatology}}
\label{fig:tc_strength}
\end{figure}

\begin{figure}[h!]
    \centering
    \includegraphics[width=0.75\textwidth]{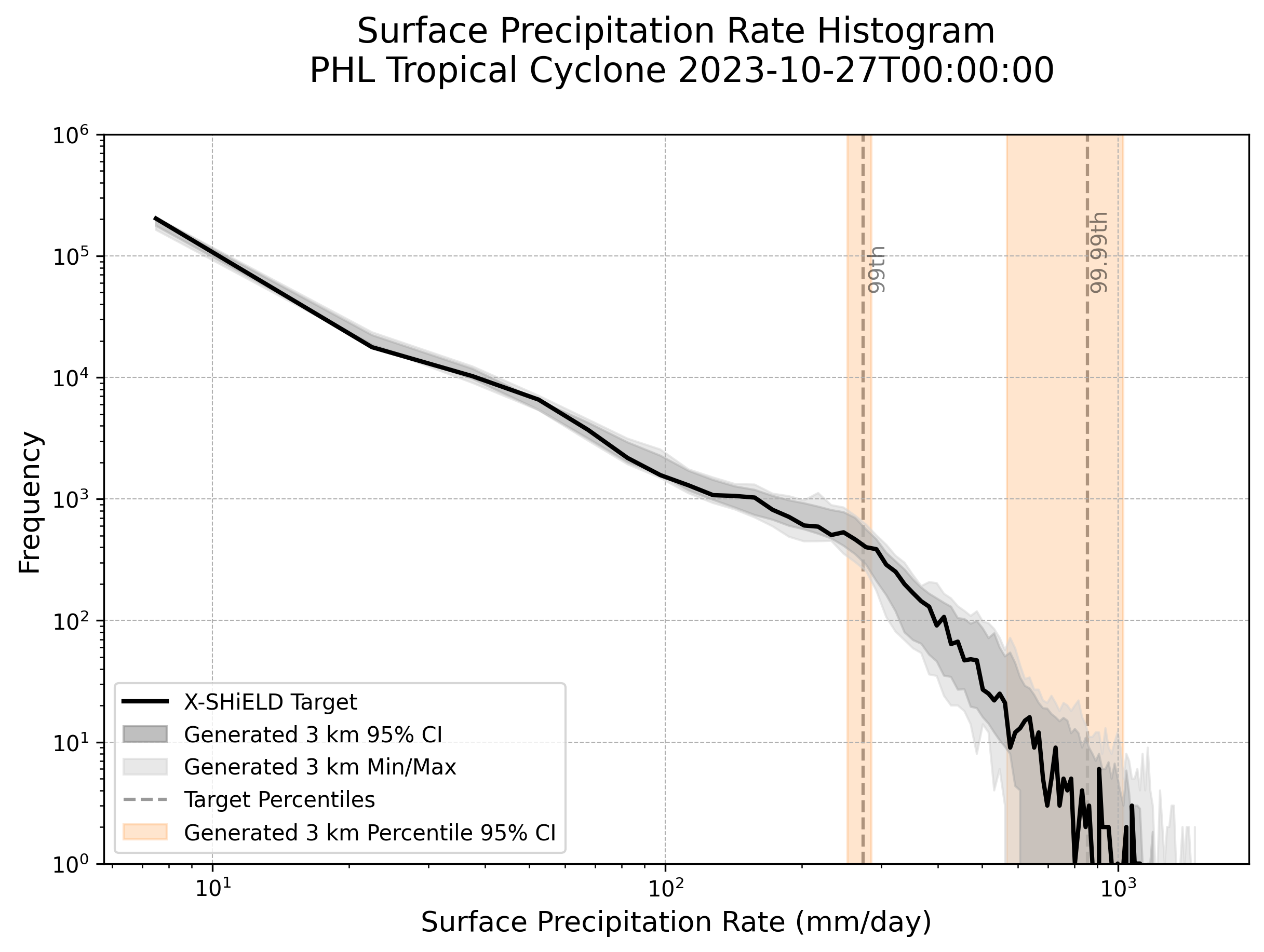}
    \caption{A comparison of histograms of a 128-member HiRO perfect prediction ensemble and of the target X-SHiELD data for 6 hr average surface precipitation rates during a simulated landfalling tropical cyclone in the Philippines (00 UTC Oct. 27 2023).  The 95\% confidence interval of the generated histogram distribution (dark grey) and the minimum and maximum histogram values (light grey) are shown.  Vertical dashed lines indicate percentiles of the target X-SHiELD data, while the orange vertical shaded region indicates the 95\% confidence interval of the same  percentiles of the HiRO generated events.}
    \label{fig:phl-hurricane-pp-hist}
\end{figure}

\begin{figure}[h!]
    \centering
    \includegraphics[width=0.9\textwidth]{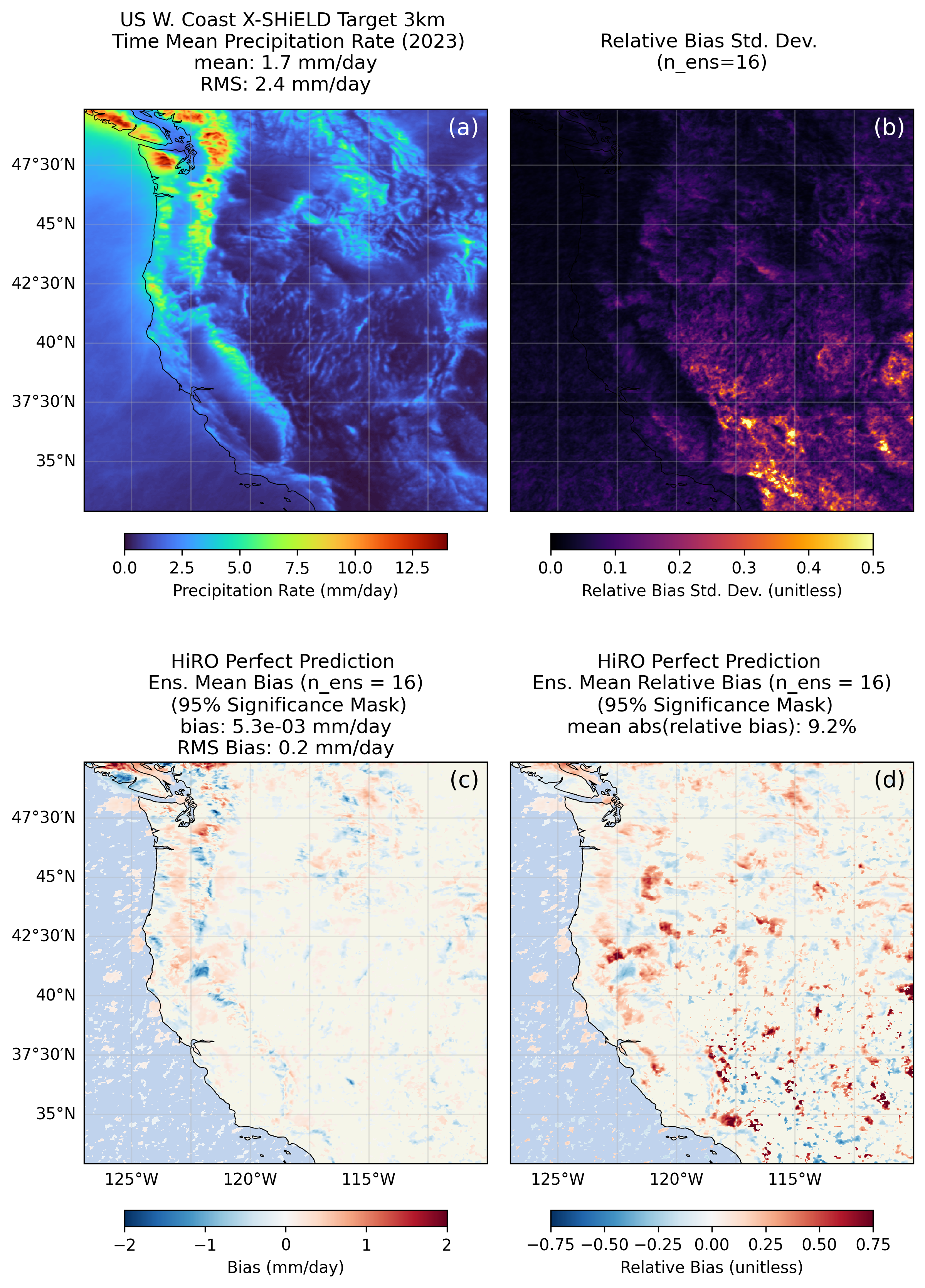}
    \caption{HiRO Perfect Prediction US West Coast ensemble mean biases w/ significance determined from the relative residuals of a 16-member perfect prediction time mean over the held out 2023 period.} 
    \label{fig:wcoast-pp-time-mean-bias-significance}
\end{figure}

\clearpage

\bibliography{references.bib}

\end{document}